\documentclass[%
 twocolumn,10pt,superscriptaddress,longbibliography,amsmath,amssymb,aps,prx,floatfix,
]{revtex4-2}

\usepackage[T1]{fontenc}

\usepackage{graphicx}
\usepackage{qcircuit}
\usepackage{mathtools}
\usepackage{amsthm}
\usepackage{tikz}
\usepackage{color}
\usepackage[colorlinks=true,linkcolor=blue,citecolor=blue, linktocpage=true,breaklinks=true]{hyperref}
\usepackage{multibib}

\usepackage{comment}

\newcommand{\tr}{\mathrm{Tr}}
 
\newtheorem{theo}{Theorem}
\newtheorem{corollary}[theo]{Corollary}

\begin{document}

\title{Conditional Entropy Production and Quantum Fluctuation Theorem of \protect\\
Dissipative Information: Theory and Experiments}

\author{Kun \surname{Zhang}}
\affiliation{Department of Chemistry, Stony Brook University, Stony Brook, New York 11794, USA}

\author{Xuanhua \surname{Wang}}
\email{These authors contributed equally to this work.}
\affiliation{Center for Theoretical Interdisciplinary Sciences, Wenzhou Institute, University of Chinese Academy of Sciences, Wenzhou, Zhejiang 325001, China}

\author{Qian \surname{Zeng}}
\email{These authors contributed equally to this work.}
\affiliation{State Key Laboratory of Electroanalytical Chemistry, Changchun Institute of Applied Chemistry, Changchun, Jilin 130022, China}

\author{Jin \surname{Wang}}
\email{jin.wang.1@stonybrook.edu}
\affiliation{Department of Chemistry, Stony Brook University, Stony Brook, New York 11794, USA}
\affiliation{Department of Physics and Astronomy, Stony Brook University, Stony Brook, New York 11794, USA}

\date{\today}

\begin{abstract}


    We study quantum conditional entropy production, {which quantifies the irreversibility of system-environment evolution from the perspective of a third system, called the reference. The reference is initially correlated with the system.} We show that the quantum unconditional entropy production with respect to the system is less than the conditional entropy production with respect to the reference, where the latter {includes a reference-induced dissipative information}. The dissipative information pinpoints the distributive correlation established between the environment and the {reference, even though they do not interact directly}. When reaching the thermal equilibrium, the system-environment evolution has a zero unconditional entropy production. However, one can still have a nonzero conditional entropy production with respect to the reference, which characterizes the informational nonequilibrium of the system-environment evolution in the view point of the reference. The additional contribution to the conditional entropy production, the dissipative information, characterizes a minimal thermodynamic cost that the system pays for maintaining the correlation with the reference. Positive dissipative information also characterizes potential work waste. {We prove that both types of entropy production and the dissipative information follow quantum fluctuation theorems when a two-point measurement is applied.} We verify the quantum fluctuation theorem for the dissipative information experimentally on IBM quantum computers. We also present examples based on the qubit collisional model and demonstrate universal nonzero dissipative information in the qubit Maxwell's demon protocol.
	
\end{abstract}

\maketitle

\section{Introduction}

Thermodynamics and quantum mechanics, as two fundamental theories, describe physics on vastly different scales but merge in the field of quantum thermodynamics \cite{vinjanampathyQuantumThermodynamics2016,deffnerQuantumThermodynamicsIntroduction2019}. The concept of entropy production, the key quantity in the second law of thermodynamics, has also been established in quantum thermodynamics \cite{landiIrreversibleEntropyProduction2021}. The irreversibility of any process is reinterpreted as the information loss to the environment \cite{espositoEntropyProductionCorrelation2010}. The stochastic version of the second law of thermodynamics, known as the fluctuation theorem \cite{evansFluctuationTheorem2002,jarzynskiEqualitiesInequalitiesIrreversibility2011,seifertStochasticThermodynamicsFluctuation2012}, has also been extended to the quantum level, commonly founded on the two-point measurement scheme \cite{tasakiJarzynskiRelationsQuantum2000,espositoNonequilibriumFluctuationsFluctuation2009,campisiColloquiumQuantumFluctuation2011,funoQuantumFluctuationTheorems2018}.


The laws of classical thermodynamics were challenged by the proposition of Maxwell's demon. An external source acquiring additional information can potentially significantly drive a system out of thermal equilibrium \cite{maruyamaColloquiumPhysicsMaxwell2009}. This consideration generated questions about the role of information in thermodynamics. Maxwell's demon is considered as a reference to the system, sometimes also called the memory, and can store the information of the system and convert it into work \cite{sagawaSecondLawThermodynamics2008,sagawaThermodynamicsInformationProcessing2012,funoQuantumNonequilibriumEqualities2015,camatiExperimentalRectificationEntropy2016,naghilooInformationGainLoss2018,tanAlternativeExperimentalWays2021}. Classical bits of the system can be partially or completely obtained after the measurement of the reference depending on the strength of the correlations between the system and the reference. In other words, if the system shares correlations with the reference, the von Neumann entropy of the system will be reduced once the information on the reference is known. This suggests that entropy and correlation are inversely related \cite{lloydUseMutualInformation1989}. The entropy of a system conditioned on the {reference} is quantified by the conditional entropy \cite{cerfNegativeEntropyInformation1997}. The conditional entropy is always upper bounded by the unconditional entropy. The two are equal if the system and the reference are uncorrelated. Negative conditional entropy suggests entanglement resources beyond classical cases \cite{horodeckiPartialQuantumInformation2005,rioThermodynamicMeaningNegative2011}.


\begin{figure}[t]
\includegraphics[width=0.8\columnwidth]{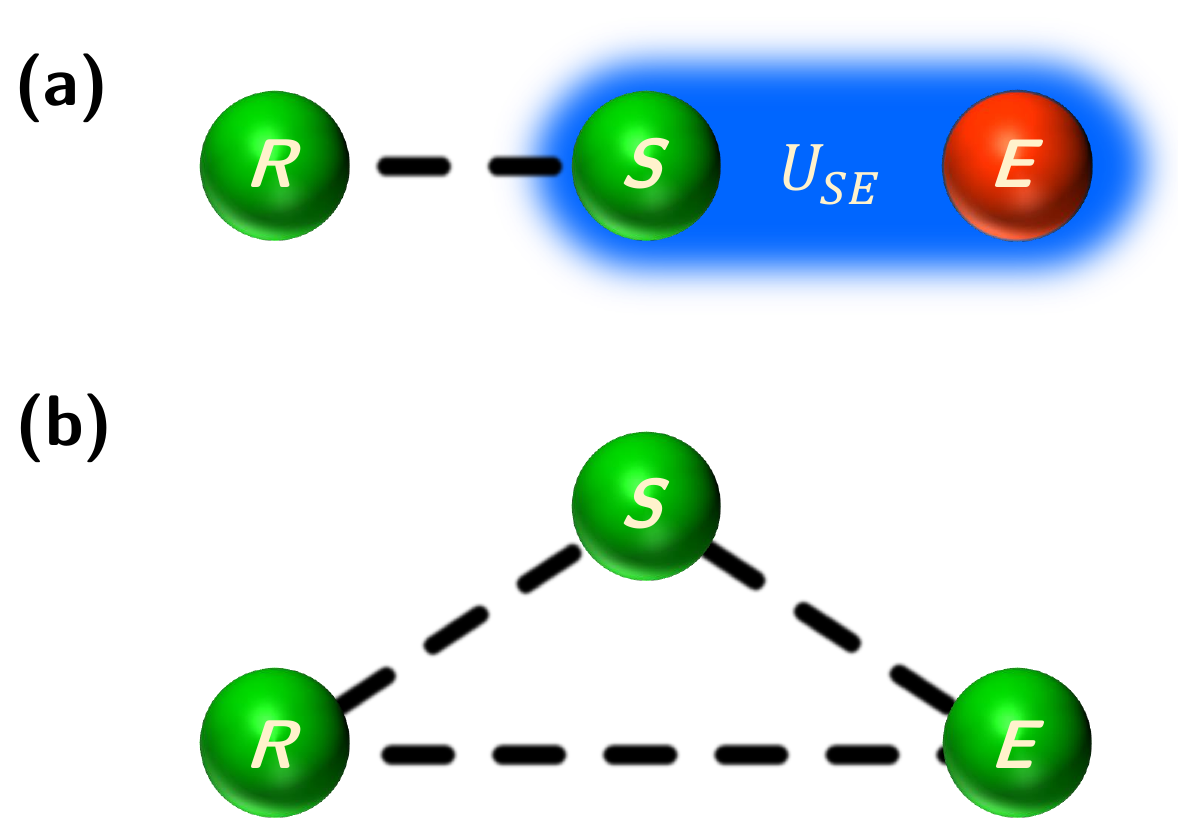}
\caption{{(a) The system S, initially correlated with the reference R, interacts with the environment E via the unitary evolution $U_{SE}$. (b) After the evolution, the reference R also establishes correlations with the environment E.}}
\label{fig_model}
\end{figure}

{In this paper, we study the conditional entropy production of a system-environment interaction, conditioned on the state of the reference. The study is based on a typical tripartite setup (system, environment, and reference --- see Fig. \ref{fig_model}). While the conditional entropy quantifies the information of the system given the information of the reference, the conditional entropy production quantifies the irreversibility of the dynamics from the viewpoint of the reference. When we say that dynamics are irreversible, we mean that we cannot retrodict the initial state of the system from the final state of the system. Irreversibility from the viewpoint of the reference means that we cannot retrodict the information between the system and the reference from the final state of the system. In a demonless setup (the bipartite setup including the system and the environment), the correlation established between the system and the environment is usually considered to be lost, and then the entropy production is related to the mutual information between the system and the environment \cite{landiIrreversibleEntropyProduction2021}. Even in setups without a reference, the information loss, entropy production, and thermodynamic laws can be conditionally defined, which is motivated by careful studies of the subjectivity of irreversibility or irretrodictability. For example, after a system-environment interaction}, if the state of the environment is known (from measurements of the environment), the information acquired with the measurements can decrease the degree of irreversibility of the dynamics \cite{belenchiaEntropyProductionContinuously2020,landiInformationalSteadyStates2022}.

When a system is coupled to the environment {to extract work}, the {reference or the demon} also establishes correlation with the environment, {even without direct interactions with the environment. {Since work can be extracted from correlations \cite{funoThermodynamicWorkGain2013,beraGeneralizedLawsThermodynamics2017,manzanoOptimalWorkExtraction2018}, the correlation established indirectly between the {reference} and the environment implies a hidden work waste}. This has been missed in previous studies on the quantum Maxwell's demon model. Our study aims to clarify such additional information loss through the study of conditional entropy production.}


In our study, we find that the conditional entropy production of the system, conditioned on the reference, is equal to or larger than the unconditional entropy production. There is a reference-induced dissipation, here called the dissipative information, which is additional to the energy dissipation of the system. As a direct consequence of the positive dissipative information, the maximum work that can be extracted from the system is always less than the conditional free energy difference, and the dissipative information is the wasted work in the work extraction. Based on a qubit collisional model \cite{landiIrreversibleEntropyProduction2021}, we demonstrate clear distinctions between thermal and informational nonequilibrium through the disparate behaviors of dissipative information, conditional entropy production, and unconditional entropy production. Furthermore, we demonstrate that the dissipative information is nonzero for any Maxwell's demon with the non-optimal feedback control, {where the hidden work waste characterized by the positive dissipative information universally exists}.

We also establish quantum fluctuation theorems associated with the conditional entropy production, based on the two-point measurement scheme. Surprisingly, we find that the dissipative information itself also follows a fluctuation theorem. This is beyond the quantum fluctuation theorems of thermodynamic quantities such as heat \cite{jarzynskiClassicalQuantumFluctuation2004} and entropy production \cite{manzanoQuantumFluctuationTheorems2018}. We test our quantum fluctuation theorem of dissipative information on IBM quantum computers. 

This paper is organized as follows. Sec. \ref{sec:conditional_ep} introduces the concept of conditional entropy production. Sec. \ref{sec:work_heat_bounds} discusses the work and heat bounds related to conditional entropy production. {Sec. \ref{sec:example_qubit_model} shows the distinct behaviors of the conditional and unconditional entropy production in a qubit collisional model. In Sec. \ref{sec:example_maxwell_demon}, we demonstrate the universal positivity of the dissipative information in the qubit Maxwell's demon model.} Sec. \ref{sec:QFT} establishes the quantum fluctuation theorems of conditional entropy production and dissipative information. {Sec. \ref{sec:ft_global} explores the quantum fluctuation theorem given by the global two-point measurement, which preserves the quantum correlation included in the conditional entropy production.} Sec. \ref{sec:experiments} presents experimental verification of the quantum fluctuation theorems on IBM quantum computers. The last section is our conclusions.

\section{Conditional entropy production}

\label{sec:conditional_ep}


{We consider a tripartite setup, including a system (S), a reference (R), and an environment (E). Initially, we assume that the system and the environment are independent, while the system and the reference are correlated. The initial state can then be denoted as $\rho_{SR}\otimes \rho_{E}$. The term reference is commonly used in quantum information science and represents the purification of the system \cite{nielsenQuantumComputationQuantum2010}. However, we do not limit ourselves to the pure state $\rho_{SR}$ in our study.}

{Next, we assume that evolution only happens between the system and the environment. In other words, the system and the reference may be spatially separated but initially share some correlations. The final state is given by
\begin{equation}
\label{eq:setup_three}
    {\rho'_{RSE}} = {U_{SE}}(\rho_{SR}\otimes{\rho_{E}}){U_{SE}^\dag}.
\end{equation}
The superscript prime identifies variables related to the final state. See Fig. \ref{fig_model} for an illustration of the above setup.}

{Since the environment interacts only with the system, the reference is unchanged during the evolution, so $\rho'_R = \rho_R$. If we trace out the reference in Eq. (\ref{eq:setup_three}), we get}
\begin{equation}
\label{eq:setup}
    {\rho'_{SE}} = {U_{SE}}(\rho_{S}\otimes{\rho_{E}}){U_{SE}^\dag}.
\end{equation}
{In other words, if we ignore the reference, we reduce the model to a bipartite model. Such a bipartite setup is commonly applied in the study of quantum thermodynamics \cite{landiIrreversibleEntropyProduction2021}.} The irreversibility of a system's evolution is justified by tracing out the environment. The (unconditional) entropy production of the system is \cite{landiIrreversibleEntropyProduction2021}
\begin{equation}
\label{def Sigma S}
    {\Sigma_{S} = \Delta \mathcal S_{S}+\Delta\Phi_{E}},
\end{equation}
where the entropy change of the system is {$\Delta \mathcal S_{S}=\mathcal S(\rho_S')-\mathcal S(\rho_S)$} and the entropy flux {$\Delta\Phi_{E}$} is from the system to the environment. {Here the entropy of the quantum state is characterized by the von Neumann entropy, i.e., $\mathcal S(\rho) = -\tr\left(\rho\ln\rho\right)$.} 


{When the degree of freedom of the environment is large, we can always approximate the initial state of the environment as thermal \cite{lindenQuantumMechanicalEvolution2009,lloydPureStateQuantum2013}, given by} ${\rho_{E}}=e^{-\beta {\mathcal H_{E}}}/{\mathcal Z_{E}}$ with the {inverse} temperature $\beta=1/T$, the Hamiltonian of the environment {$\mathcal H_{E}$, and the corresponding partition function $\mathcal Z_E$}. The Boltzmann constant is set to 1. The entropy flux is related to the heat flux ${\Delta\Phi_{E}} = \beta {Q_{E}}$, where the heat is defined as the energy change of the environment ${Q_{E}} = \tr(({\rho_{E}'}-{\rho_{E}}){\mathcal H_{E}})$ \cite{landiIrreversibleEntropyProduction2021}. Local reversible evolutions, such as ${U_{SE}}= U_{S}\otimes {1\!\!1_{E}}$, give $\Sigma_{S}=0$. Correlation with the environment indicates irreversibility of the dynamics \cite{espositoEntropyProductionCorrelation2010}. 


{If we restore to our tripartite setup including the reference, we can consider the entropy production of the combined system and reference, which is given by
\begin{equation}
    \Sigma_{SR} = {\Delta \mathcal S_{SR}}+{\Delta\Phi_{E}},
\end{equation}
with the entropy change $\Delta \mathcal S_{SR} = \mathcal S(\rho'_{SR})-\mathcal S(\rho_{SR})$ and the entropy flux $\Delta\Phi_{E}$. Although the state of reference is unchanged, we have $\Delta \mathcal S_{SR}\neq \Delta \mathcal S_{S}$ in general cases. The physical meaning of $\Sigma_{SR}\neq \Sigma_{S}$ seems unclear since the reference has no interaction with the environment. Instead of the total entropy production $\Sigma_{SR}$, we consider the conditional entropy production, which characterizes the irreversibility of the dynamics from the viewpoint of the {reference (or demon)} rather than the system itself.} The conditional entropy production of the system conditioned on the {reference} is defined as
\begin{equation}
\label{def SM}
    {\Sigma_{S|R}} = {\Delta \mathcal S_{S|R}}+{\Delta\Phi_{E|R}},
\end{equation}
with the conditional entropy change ${\Delta \mathcal S_{S|R}} = {\mathcal S'_{S|R}}-{\mathcal S_{S|R}}$ and the conditional entropy flux {$\Delta\Phi_{E|R}$}. The conditional entropy is formulated as ${\mathcal S_{S|R}} = {\mathcal S(\rho_{SR})}-{\mathcal S(\rho_{R})}$, quantifying the uncertainty remaining in the system given the information of the {reference}.

{There may appear to be ambiguity in the definition of the conditional entropy flux, which quantifies the amount of entropy flowing into the environment. Under the assumption that the environment is thermal with a well-defined temperature, the conditional entropy flux is given by the conditional heat flux. In quantum thermodynamics, the definitions of heat and work are still controversial because the boundary between work and heat is vague. One way to resolve this issue is to define the heat independently as the energy change of the environment, $Q_{E} = \tr((\rho_{E}'-\rho_{E})\mathcal H_{E})$. From such a viewpoint, the conditional entropy flux is equal to the unconditional entropy flux,
\begin{equation}
\label{eq:entropy_flux_equal}
    {\Delta\Phi_{E|R} = \Delta\Phi_{E},}
\end{equation}
since knowing the state of the reference does not alter the energy change of the environment. Similar arguments can be applied if the heat is defined by the entropy change of the environment \cite{beraGeneralizedLawsThermodynamics2017}. The conditional entropy flux is also equal to the unconditional entropy flux when the conditional information is in terms of the environment \cite{landiInformationalSteadyStates2022}. It is only when the system-environment evolution depends on the state of the reference then the magnitude of the conditional entropy flux is different from the unconditional entropy flux. We leave this scenario for future study.}

{It is easy to show that
\begin{equation}
    \Delta\mathcal S_{S|R} = \Delta\mathcal S_{S|R}-\Delta\mathcal S_R = \Delta\mathcal S_{SR}.
\end{equation}
Combining this with the condition in Eq. (\ref{eq:entropy_flux_equal}), we can conclude that
$$
\Sigma_{SR} = \Sigma_{S|R}.
$$
Now we can interpret the total entropy production $\Sigma_{SR}$ as the conditional entropy production $\Sigma_{S|R}$. The total entropy production $\Sigma_{SR}$ treats the system and the reference as being on equal footing even though the interaction $U_{SE}$ is only applied to the system. The role of the reference in $\Sigma_{S|R}$ is much clearer than its role in $\Sigma_{SR}$, which motivates us to study the conditional entropy rather than the total entropy production.}

{Both the unconditional entropy production $\Sigma_S$ and the conditional entropy production $\Sigma_{S|R}$ are positively defined. Moreover, we have}
\begin{theo}
\label{theorem 1}
\normalfont${\Sigma_{S|R}}\geq\Sigma_{S}\geq0$.
\end{theo}
\begin{proof}
The positivity of the entropy production $\Sigma_S$ is guaranteed by the positivity of the relative entropy \cite{landiIrreversibleEntropyProduction2021}. {The mismatch between the conditional and unconditional entropy production is given by the difference between the conditional and unconditional entropy changes, i.e.,
\begin{equation}
    \Sigma_{S|R} - \Sigma_S = \Delta S_{S|R} - \Delta S_S,
\end{equation}
which gives
\begin{equation}
\label{eq:mismatch_expansion}
    \Sigma_{S|R} - \Sigma_S = \mathcal S(\rho'_{SR}) - \mathcal S(\rho_{SR}) - \mathcal S(\rho'_S) + \mathcal S(\rho_S).
\end{equation}
Consider the mutual information, defined by
\begin{equation}
    \mathcal I_{S;R} = \mathcal S(\rho_S)+\mathcal S(\rho_R)-\mathcal S(\rho_{SR}),
\end{equation}
which quantifies the degree of correlation between the two parties \cite{wildeClassicalQuantumShannon2017}. The final state mutual information is denoted as $\mathcal I'_{S;R}$. By substituting the definition of the mutual information into Eq. (\ref{eq:mismatch_expansion}), one sees that the mismatch can be rewritten as a mutual information change, i.e.,
\begin{equation}
\label{eq_s1}
    \Sigma_{S|R} - \Sigma_S = \mathcal I_{S;R} - \mathcal I'_{S;R}.
\end{equation}
}

{Unitary evolution preserves the information. The initial correlation between the system and the environment is preserved as $\mathcal I_{S;R} = \mathcal I'_{SE:R}$. In other words, the information between the system and the reference spreads to the environment because of the evolution between the system and the environment.} We can then rewrite Eq. (\ref{eq_s1}) as
\begin{equation}
\label{eq_s4}
    {\Sigma_{S|R}} - \Sigma_{S}  = {\mathcal I'_{SE:R}} - {\mathcal I'_{S;R}} \geq 0.
\end{equation}
The last inequality comes from the monotonicity of the quantum mutual information \cite{wildeClassicalQuantumShannon2017}, {also called the quantum data-processing inequality \cite{schumacherQuantumDataProcessing1996}. It can be derived from} the strong subadditivity of the von Neumann entropy \cite{liebProofStrongSubadditivity1973}.
\end{proof}

The {reference} stores the information about the system. The system-environment evolution corrupts the correlation between the system and the {reference}. In other words, the {reference} is less related to the state of the system after the system's evolution with the environment. It happens even when the system's reduced density matrix is unchanged (in equilibrium with the environment). To emphasize the difference between unconditional and conditional entropy production, we define their mismatch as
\begin{equation}
\label{eq SM S I}
    \Sigma_{I} = {\Sigma_{S|R}} - \Sigma_{S}\,,
\end{equation}
which is called the dissipative information. {This quantity first appeared in a study on the classical Maxwell's demon model \cite{zengNewFluctuationTheorems}.} Based on Theorem \ref{theorem 1}, we know that
\begin{corollary}
\label{corollary_2}
\normalfont$\Sigma_{I} \geq 0$.
\end{corollary}



On the one hand, the unconditional entropy production is the lower bound of the conditional entropy production, i.e.,
\begin{equation}
    {\Sigma_{S|R}}\geq\Sigma_{S}\,.
\end{equation}
An equals sign applies when the environment only {interacts with} the subspace of the system, while only the complementary subspace {initially correlates} to the {reference} \cite{haydenStructureStatesWhich2004}. On the other hand, the lower bound, which is expressed in terms of the dissipative information, is given by
\begin{equation}
    {\Sigma_{S|R}}\geq\Sigma_{I}\,,
\end{equation}
and is reached when the system and the environment are at a global fixed point in the evolution ${U_{SE}}$. In other words, the system is in thermal equilibrium with the environment while the dissipative information is still nonzero. {Since $\Sigma_I$ characterizes a purely informational dissipation, we call it informational nonequilibrium when $\Sigma_I>0$, which is independent of the thermal equilibrium or nonequilibrium. See Sec. \ref{sec:example_qubit_model} for more examples.}

Another way to understand the dissipative information is to keep track of information flow. More specifically, the dissipated information between the system and the reference becomes the correlation between the reference and the environment, even though the reference and the environment do not directly interact. This is called the distributive correlation \cite{cubittSeparableStatesCan2003,chuanQuantumDiscordBounds2012} (see Fig. \ref{fig_model}). Quantitatively, we have
\begin{theo}
\label{theorem 2}
\normalfont$\Sigma_{I}={\mathcal I'_{{R};{E}|{S}}}$.
\end{theo}
\noindent Here ${\mathcal I'_{{R};{E}|{S}}}$ is the conditional mutual information of the final state between the environment and the {reference} (conditioned on the system). It is defined as ${\mathcal I_{{R};{E}|{S}}} = {\mathcal S(\rho_{SE})} + {\mathcal S(\rho_{SR})}- \mathcal S(\rho_{S})-{\mathcal S(\rho_{RSE})}$.
\begin{proof}
The conditional mutual information ${\mathcal I'_{{R};{E}|{S}}}$ can be rewritten as
\begin{equation}
    {\mathcal I'_{{R};{E}|{S}}} =  {\mathcal I'_{SE:R}} - {\mathcal I'_{S;R}}.
\end{equation}
Then, according to Eq. (\ref{eq_s4}), we have
\begin{equation}
    \Sigma_{I} = {\mathcal I'_{{R};{E}|{S}}}.
\end{equation}
\end{proof}

The information related to the environment is considered to be lost since we do not have access to the environment in general. The conditional mutual information ${\mathcal I'_{{R};{E}|{S}}}$ should also be identified as an entropy production, but{ it is not included in the unconditional entropy production.} Previous studies have revealed that the conditional mutual information ${\mathcal I_{{R};{E}|{S}}}$ bounds the fidelity to reconstruct the combined system (the system plus the {reference}) from the system alone \cite{fawziQuantumConditionalMutual2015,brandaoQuantumConditionalMutual2015}. {The dissipative information, understood as the conditional mutual information, directly reflects the irretrodictability of the correlation between the system and the reference, which corresponds to the irreversibility of the dynamics from the viewpoint of the reference.}


{If the reference is interpreted as Maxwell's demon, we can view the system-environment evolution as a work extraction process, which occurs after the demon's control. The essence of Maxwell's demon is to acquire the state of the system, and the entropy of the system can then be reduced. The work extraction process can be designed independently of the state of the demon. Therefore, the work-extracting evolution is limited between the system and the environment, and this is the case in our study. More discussion on the dissipative information and Maxwell's demon model can be found in Sec. \ref{sec:example_maxwell_demon}.} The superficial negative entropy production in Maxwell's demon model is due to the feedback control protocols that make the system open to a third party \cite{sagawaSecondLawThermodynamics2008,sagawaThermodynamicsInformationProcessing2012,funoQuantumNonequilibriumEqualities2015}. Our arguments on conditional entropy production {clarify} that the entropy production of a closed system (system plus environment) conditioned on the third party is always positive.

\section{Work and heat bounds}

\label{sec:work_heat_bounds}

{A direct consequence of positive dissipative information is that the bounds of the work and heat given by the unconditional entropy production are tighter than those given by the conditional entropy production.} To see this, consider the evolution {$U_{SE}$} as a work extraction protocol, as was proposed in \cite{skrzypczykWorkExtractionThermodynamics2014}. The maximal amount of work extractable from the system is bounded by the change of the nonequilibrium free energy {$\Delta \mathcal F_{S} = \mathcal F'_{S} - \mathcal F_{S}$}, with {$\mathcal F_{S} = \tr(\mathcal H_{S}\rho_{S})-T\mathcal S(\rho_{S})$} \cite{parrondoThermodynamicsInformation2015}. {The Hamiltonian of the system is time-dependent. Therefore, we may have $\mathcal H'_S\neq \mathcal H_S$. For specific examples, see Sec. \ref{sec:example_qubit_model}.} The ordinary work and heat bounds are given by the nonnegativity of the unconditional entropy production $\Sigma_S$ (\ref{def Sigma S}). Specifically, the work done on the system satisfies $W_\text{ext}\geq \Delta \mathcal F_{S}$. The associated heat is bounded by $Q_{S}\leq T\Delta \mathcal S_{S}$, where $Q_{S}$ is the heat flow from the environment to the system and $Q_{S}=-{Q_{E}}$. Note that extracting the work means $W_\text{ext}< 0$ and $Q_{S}>0$. 

The conditional entropy production bounds the work in terms of the change of the conditional free energy: $W_\text{ext}\geq {\Delta \mathcal F_{S|R}}$. {Here we define the conditional free energy as}
\begin{equation}
    {\mathcal F_{S|R} = \tr(\mathcal H_{S}\rho_{S})-T\mathcal S_{S|R}.}
\end{equation}
The conditional free energy is equal to the maximal extractable work from the system given the information of the {reference}. In other words, more work can be extracted from the correlation between the system and the {reference} if such a correlation is accessible \cite{funoThermodynamicWorkGain2013,manzanoOptimalWorkExtraction2018,beraGeneralizedLawsThermodynamics2017}. 

However, we can see that the work bound $W_\text{ext}\geq {\Delta \mathcal F_{S|R}}$ is not tight compared to the more common or conventional bound $W_\text{ext}\geq \Delta \mathcal F_{S}$, since
\begin{equation}
\label{work bound}
    {W_\text{ext}\geq \Delta \mathcal F_{S}=\Delta \mathcal F_{S|R} +T\Sigma_{I}\geq \Delta \mathcal F_{S|R},}
\end{equation}
where $\Sigma_{I}\geq 0$ according to Corollary \ref{corollary_2}. The corresponding heat bound is
\begin{equation}
\label{heat bound}
{Q_S\leq T\Delta \mathcal S_{S}=T\Delta\mathcal S_{S|R} -T\Sigma_I\leq T\Delta \mathcal S_{S|R}}.
\end{equation}
Since the working protocol {$U_{SE}$} only locally interacts with the system, the information stored in the {reference} is not applied. While the correlation between the system and the {reference} is corrupted proportionally to $\Sigma_{I}$, the amount of correlation work $T\Sigma_{I}$ is wasted. The ``wasted'' work is stored as the correlation between the {reference} and the environment, as suggested by Theorem \ref{theorem 2}. According to the protocol in \cite{beraGeneralizedLawsThermodynamics2017}, such a correlation can be converted into work {if the joint state of the environment and the reference is accessible}.

\section{Examples for the qubit model with an isothermal process}

\label{sec:example_qubit_model}

{In this section, we demonstrate the different behaviors of the unconditional and conditional entropy production based on concrete examples. We consider the isothermal process based on the qubits collisional model \cite{landiIrreversibleEntropyProduction2021}.} Consider the initial thermal states, given by ${\rho_{S(R)}}=e^{-{\beta_{S(R)}\mathcal H_{S(R)}}}/{\mathcal Z_{S(R)}}$, with an initial Hamiltonian ${\mathcal H_{S(R)}} = {E_{S(R)}}{|1\rangle_{S(R)}\langle1|}$, an effective {inverse} temperature ${\beta_{S(R)}}$, and a corresponding partition function ${\mathcal Z_{S(R)}}$. We assume $E_{S} = {E_{R}}$ {and $\beta_{S} = \beta_{R}$} for simplicity. The correlation can be categorized as classical or quantum. One example of classical correlation is 
\begin{equation}
\label{eq:rho_sr^c}
    {\rho_{SR}^{c}} = p{|00\rangle_{SR}\langle 00|}+(1-p){|11\rangle_{SR}\langle 11|},
\end{equation}
with $p = (1+e^{-\beta_{S} E_S})^{-1}$. One example of quantum correlation, given by 
\begin{equation}
\label{eq:rho_sr^q}
    {\rho_{SR}^{q}} = {|\psi_{SR}\rangle\langle\psi_{SR}|},
\end{equation}
with ${|\psi_{SR}\rangle} = \sqrt{p}{|00\rangle_{SR}}+\sqrt{1-p}{|11\rangle_{SR}}$, characterizes the maximal quantum correlation between the system and the {reference}, {i.e., the purification \cite{nielsenQuantumComputationQuantum2010}}. Quantum entanglement gives rise to the negative conditional entropy ${\mathcal S_{{S}|{R}}} = -\mathcal S(\rho_S)$ \cite{horodeckiPartialQuantumInformation2005,rioThermodynamicMeaningNegative2011}.

\begin{figure}[t]
\includegraphics[width=1\columnwidth]{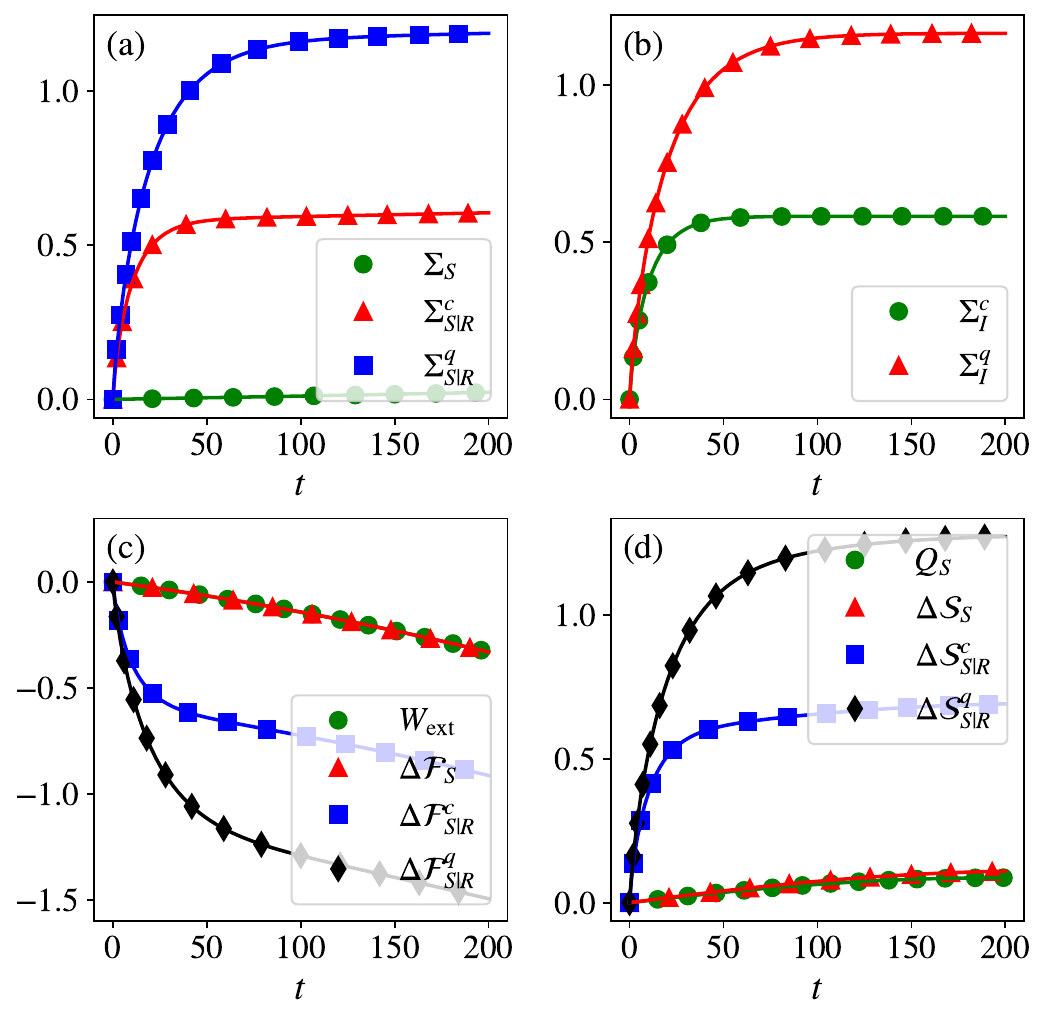}
\caption{The dynamics of (a) the entropy production and (b) the dissipative information; (c) a comparison between the work, the change in free energy, and the conditional free energy in terms of the evolution time; (d) a comparison between the heat, the change in entropy, and the conditional entropy in terms of the evolution time. The parameters are set as {$\beta E_S = 10\beta E'_S=1$}. At each step, the system quenches the energy by $\delta E=4.5\times 10^{-3}E_S$. The coupling strength between the system and the environment is $g=0.1$.}
\label{fig_example}
\end{figure}

We consider a quasistatic isothermal process \cite{skrzypczykWorkExtractionThermodynamics2014}, where the system is doing work while absorbing heat from the environment. The environment consists of thermal qubits at the {inverse} temperature $\beta$. At each step, the system quenches the Hamiltonian by a value $\delta E$, then interacts with a qubit from the environment via the XY interaction ${U_{{SE}}}$, given by
{
\begin{equation}
\label{eq:XY_interaction}
    {U_{{SE}}}=e^{-ig(\sigma^x_{S}{\sigma^y_{{E}}}-\sigma^y_{S}{\sigma^x_{{E}}})},
\end{equation} 
with $g$ the coupling parameter. The Planck constant is set to 1, and $\sigma^{x}$ and $\sigma^{y}$ are the Pauli matrices. Such an interaction has been experimentally realized in nuclear spin systems \cite{micadeiReversingDirectionHeat2019}. The quantum operation on the system S, given by
\begin{equation}
    \mathcal E(\rho_{S}) = {\tr_{{E}}}\left({U_{{SE}}}(\rho_{S}\otimes {\rho_{{E}}}){U^\dag_{{SE}}}\right),
\end{equation}
is a thermalization channel, also called the generalized amplitude damping channel \cite{nielsenQuantumComputationQuantum2010}.}

We quench the energy of state $|1\rangle_{S}$ in each step in order to extract work from the system. Suppose that at step $t$, the energy of state $|1\rangle_{S}$ is $E_{S}^{(t)}$ {and the energy level of the environmental qubit is $E^{(t)}_\text{Env}$. After the XY interaction $U_{SE}$ given by Eq. (\ref{eq:XY_interaction}), the system reaches a new thermal state with the energy of state $|1\rangle_S$ becoming $E_{S}^{(t)}-\delta E$ (a quasistatic process). Here the quenched energy $\delta E$ is determined by the relation
}
\begin{multline}
    {E^{(t)}_\text{Env}} = \\
    \frac 1 \beta \ln\left(\frac{\left(\cos^2(2g)e^{\beta \delta E}-\sin^2(2g)e^{\beta E_{S}^{(t)}}-1\right)e^{\beta E_{S}^{(t)}}}{\cos^2(2g)e^{\beta E_{S}^{(t)}}-\sin^2(2g)e^{\beta \delta E}-e^{\beta (E_{S}^{(t)}+\delta E)}}\right).
\end{multline}
When $g=\pi/4$, the operation ${U_{{SE}}}$ performs a swap-like operation {and the quenched energy follows the relation $\delta E = E_{S}^{(t)} - {E_\text{Env}^{(t)}}$.} Note that ${U_{{SE}}}$ is a strict energy conservation operation as $\delta E\rightarrow 0$.


The quasistatic limit $\delta E\rightarrow 0$ leads to $\Sigma_{S}\rightarrow 0$ (see the numerical results with small values of $\delta E$ in Fig. \ref{fig_example}). However, the conditional entropy production is not zero. {In particular, we have ${\Sigma_{S|R}^{q}}>{\Sigma_{S|R}^{c}}>\Sigma_{S}$.} The superscript $c$ or $q$ denotes a classical or quantum correlation between the system and the {reference} at the initial time. {This example is designed to be in thermal equilibrium at each step. However, the initial correlation between the system and the reference suggests an informational nonequilibrium state, which dissipates at each step.} The dissipative information given by the quantum correlation is larger than that given by the classical correlation, $\Sigma_{I}^{q}>\Sigma_{I}^{c}$, because the quantum correlation is stronger than the classical correlation. After $t>100$, the system and the reference are uncorrelated. Then the informational equilibrium has been reached. The dissipative information reaches a maximum with $\Sigma_{I}^{q} = 2\Sigma_{I}^{c} = 2\mathcal S(\rho_{S})$, which also represents the maximal quantum and classical correlations established between the {reference} and the environment.

At each step, the work done on the system (via quenching) is equal to
\begin{equation}
    W^{(t)}_\text{ext} = -\delta E \frac{e^{-\beta(E_S-t\delta E)}}{1+e^{-\beta(E_S-t\delta E)}}.
\end{equation}
At the limit $\delta E\rightarrow 0$, the total work is
\begin{equation}
    W_\text{ext} = \lim_{\delta E\rightarrow 0}\sum_t W^{(t)}_\text{ext} = \frac 1 \beta \ln\left(\frac{1+e^{-\beta E_S}}{1+e^{-\beta E'_S}}\right),
\end{equation}
{with the final energy of the system $E_S'$.} The work saturates the change of the free energy given by
\begin{equation}
    \Delta \mathcal F_{S} = \mathcal F'_{S} - \mathcal F_{S} = \frac 1 \beta \ln\left(\frac{1+e^{-\beta E_S}}{1+e^{-\beta E'_S}}\right).
\end{equation}
However, the change in the conditional free energy and the change in the conditional entropy suggest work waste according to the inequalities in Eqs. (\ref{work bound}) and (\ref{heat bound}). See Fig. \ref{fig_example} for the numerical results.

\section{Universal nonzero dissipative information in the qubit Maxwell's demon}

\label{sec:example_maxwell_demon}

\begin{figure}[t]
\centering
\includegraphics[width=\columnwidth]{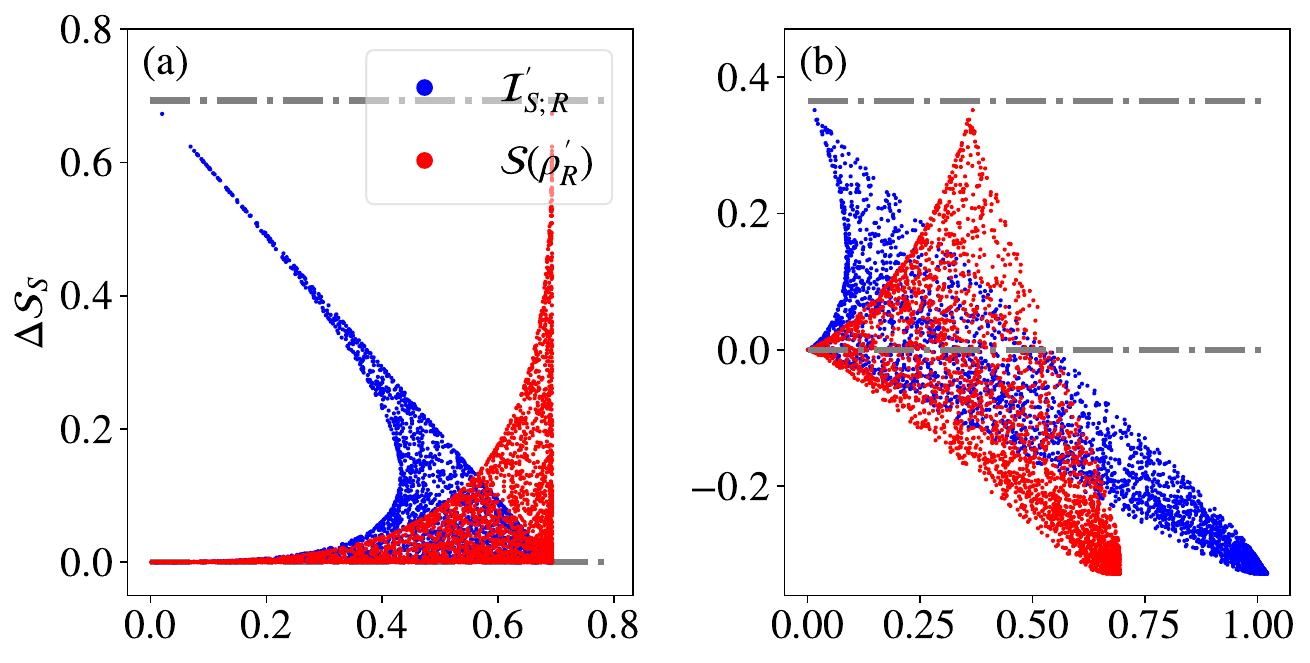}
\caption{The entropy change of the system in terms of the
final state mutual information (between the system and the {reference}) and the entropy of the {reference}, given by the random feedback control operations performed on the system and the {reference}. The {reference} has a pure initial state. The system is initially in a thermal state with {$\rho_{S} = \exp(-\beta_{S}E_S|1\rangle_{S}\langle 1|)/\mathcal Z_{S}$}. The system has the initial parameters (a) $\beta_{S} = 0$ and (b) $\beta_{S}E_S = 2$. The top
gray line is the maximum of $\Delta \mathcal S_{S}$. The bottom gray line is the threshold at which $\Delta \mathcal S_{S}=0$.}
\label{fig_demon1}
\end{figure}

{The typical tripartite setup is commonly applied to the study of Maxwell's demon \cite{sagawaSecondLawThermodynamics2008,sagawaThermodynamicsInformationProcessing2012,funoQuantumNonequilibriumEqualities2015,camatiExperimentalRectificationEntropy2016,naghilooInformationGainLoss2018,tanAlternativeExperimentalWays2021}. The demon acquires the information of the system and stores it in the reference, also called the memory. Then, based on the information of the reference, the demon controls the system and reduces its entropy. After the entropy of the system is reduced, work can be extracted from the system. The above process is called the feedback control operation and is the essence of Maxwell's demon. In the following, we consider the qubit Maxwell's demon model, where both the system and the reference (memory) are one qubit. }


Suppose that the correlation between the reference and the system is established by the unitary operation $U_{SR}$. The feedback control operation can be measurement-type or unitary-type. The unitary-type operation can be realized with two-qubit controlled gates:
\begin{align}
    \Lambda^{(0)} =& {|0\rangle_{R}\langle 0|}\otimes U^{(0)}_{S} +  {|1\rangle_{R}\langle 1|}\otimes 1\!\!1_{S},\\
    \Lambda^{(1)} =& {|0\rangle_{R}\langle 0|}\otimes 1\!\!1_{S} +  {|1\rangle_{R}\langle 1|}\otimes U^{(1)}_{S}.
\end{align}
The combined operation then performed by the demon is ${\tilde U_{SR} = \Lambda^{(1)} \Lambda^{(0)} {U_{SR}}}$. 

To examine the optimal operation ${\tilde U_{SR}}$, we randomly choose the two-qubit gates ${\tilde U_{SR}}$ acting on {SR}. The random two-qubit gates are given by three randomly chosen non-local parameters \cite{krausOptimalCreationEntanglement2001}. Note that local unitary evolutions do not change the entropy of the system. If the {inverse} temperature of the system is $\beta_{S} = 0$, the initial state of the system is a completely mixed state. The demon's operation ${\tilde U_{SR}}$ can only decrease the entropy of the system. However, only when we have a swap gate (up to some single-qubit gates) \cite{nielsenQuantumComputationQuantum2010}, the entropy of the system reduces to zero, assuming that the initial state of the reference is pure. {If $\tilde U_{SR}$ is not a swap-like operation, the system and the reference (memory) remain correlated after the evolution $\tilde U_{SR}$.} See Fig. \ref{fig_demon1} for the numerical results. When the initial state of the system is not completely mixed, such as $\beta_{S}E_S = 2$, the demon's operation can either increase or decrease the entropy of the system. The increased entropy of the system comes from the correlation with the {demon's memory}. {For the qubit Maxwell's demon, the optimal operation ${\tilde U_{SR}}$ is always the swap-like gate,} which reduces the entropy of the system to zero.

{For the measurement-type Maxwell's demon, the demon's control operation is given by a conditional local operation based on the measurement performed on the reference (memory).} In theoretical analyses, we can postpone the measurement and the measurement-type demon becomes equivalent to the unitary-type demon with additional measurements performed at the end. In other words, we have
\begin{equation}
\Qcircuit @C=0.9em @R=1.5em {
& \ctrlo{1} \qw & \ctrl{1} \qw & \meter &&&&  \meter \qw \cwx[1]\\
& \gate{U^{(0)}} & \gate{U^{(1)}} & \qw &\raisebox{0.8cm}{~~~=}&&& \gate{U^{(k)}} & \qw }
\end{equation}
with $k=0,1$ as the measurement results. To minimize the back actions from the measurement, we assume that the measurement is performed on the eigenbasis of the reference (denoted as $|k\rangle_R$). Measurements performed on the non-eigenbasis may introduce more randomness during the protocol \cite{naghilooInformationGainLoss2018}. 

\begin{figure}[t]
\centering
\includegraphics[width=\columnwidth]{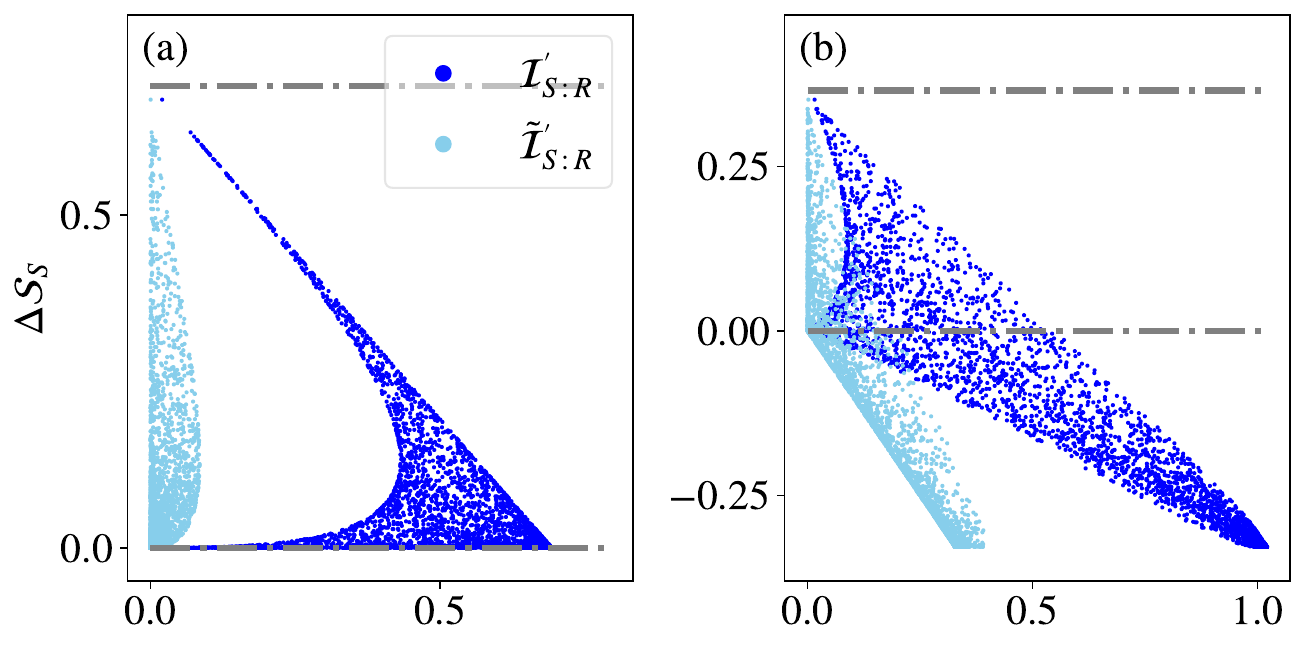}
\caption{The mutual information between the system and the {reference} after the random feedback control operations. The initial state of system is a thermal state with (a) $\beta_
{S}=0$ and (b) $\beta_{S}E_S=2$. The mutual information ${\tilde{\mathcal I}'_{S;R}}$, given by Eq. (\ref{def:tilde_I}), quantifies the classical correlation. The top gray line is the maximum of $\Delta\mathcal S_{S}$. The bottom gray line is the threshold at which $\Delta\mathcal S_{S}=0$. }
\label{fig_demon2}
\end{figure}

The difference between the unitary feedback control operation and the feedback control operation with the measurement is the type of correlation (classical or quantum) between the system and the {reference} after the operation. Fig. \ref{fig_demon1} shows that there is some remaining correlation if the demon does not perform the optimal swap-like operation. {One can check that the measurement-type demon can result in some remaining classical correlation between the system and the reference (see Fig. \ref{fig_demon2}). }The mutual information ${\tilde{\mathcal I}_{S;R}}$ means that the {reference} is dephased {because of the measurement}, i.e.,
\begin{equation}
\label{def:tilde_I}
    {\tilde{\mathcal I}_{S;R}} = \mathcal S(\rho_{S})+{\mathcal S(\rho_{R})}-{\mathcal S(\tilde{\rho}_{SR})},
\end{equation}
with
\begin{equation}
    {\tilde{\rho}_{SR}} = \sum_k |k\rangle_R\langle k| {\rho_{SR}}|k\rangle_R\langle k|.
\end{equation}
Here $|k\rangle_R$ is the eigenbasis of the {reference}. Unlike the unitary feedback control operations, non-optimal feedback control with the measurement can leave the system and the {reference} uncorrelated. However, this is not guaranteed.


{The simulation results in Figs. \ref{fig_demon1} and \ref{fig_demon2} show a universal correlation between the system and the reference (memory) after the demon's feedback control operations, either for the unitary-type or the measurement-type demon. When we couple the system to the environment to extract work, such as in the protocol shown in Sec. \ref{sec:example_qubit_model}, the nonzero dissipative information is universal.} For example, suppose that the system and the {reference} have a classically correlated initial state (before the work extraction)
\begin{equation}
\label{eq:rho_sr^c_noisy}
    {\rho_{SR}^{c}(\varepsilon_{c})} = {\rho_{SR}^{c}} - \varepsilon_{c}p(1-p)(\sigma_{S}^z\otimes{\sigma_{R}^z}),
\end{equation}
or an initial state with the quantum correlation
\begin{equation}
\label{eq:rho_sr^q_noisy}
    {\rho_{SR}^{q}(\varepsilon_{q})} = {\rho_{SR}^{q}} - \varepsilon_{q}\sqrt{p(1-p)}({|00\rangle_{SR}\langle11|}+{|11\rangle_{SR}\langle00|}),
\end{equation}
with the noise parameters $\varepsilon_{c}, \varepsilon_{q}\in[0,1]$. {Here ${\rho_{SR}^{c}}$ and ${\rho_{SR}^{q}}$ are given by Eqs. (\ref{eq:rho_sr^c}) and (\ref{eq:rho_sr^q}), respectively.} Note that ${\rho_{SR}^{c}(1)} = \rho_{S}\otimes {\rho_{R}}$ and ${\rho_{SR}^{q}(1)}={\rho_{SR}^{c}(0)}$. 

\begin{figure}[t]
\includegraphics[width=1\columnwidth]{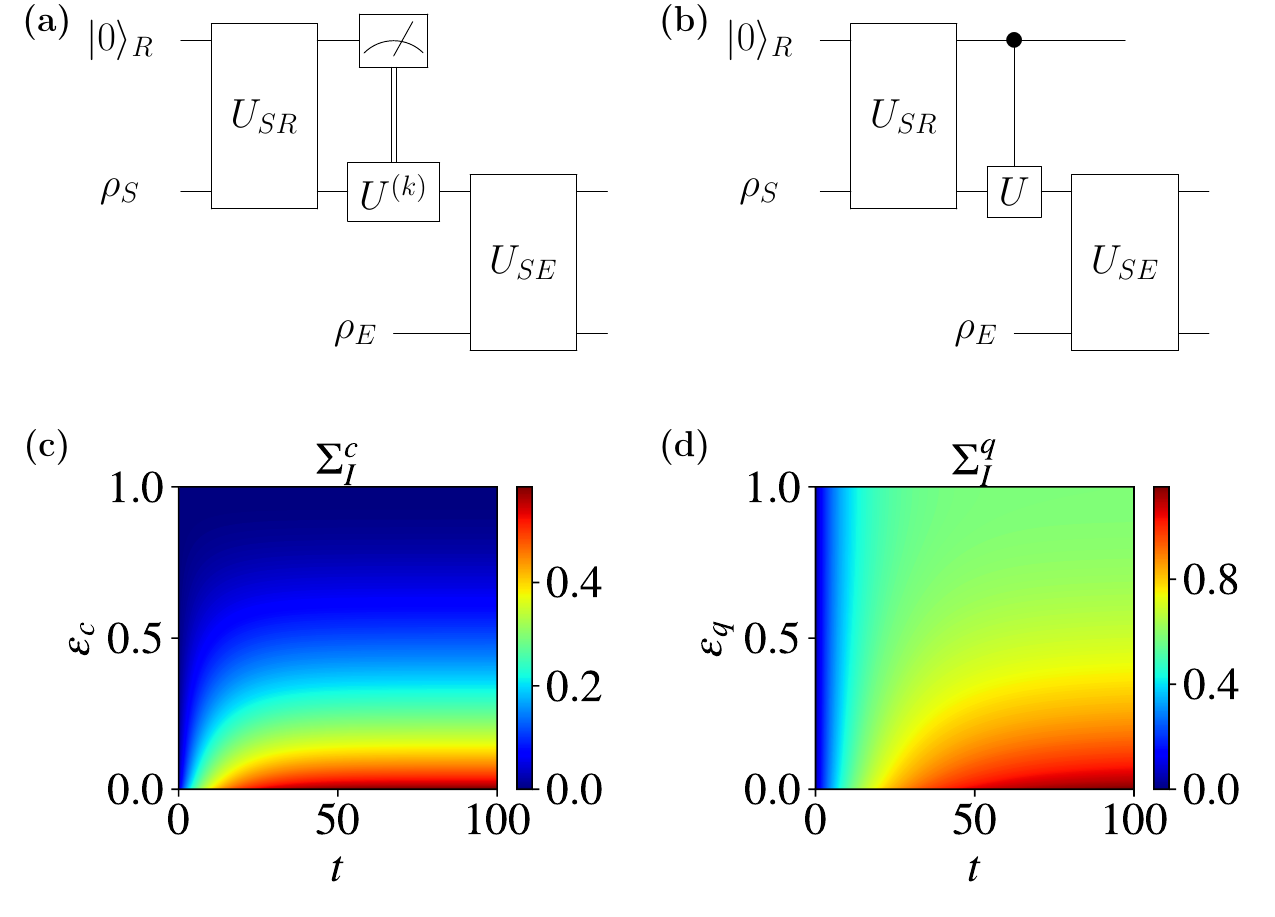}
\caption{(a) The measurement-type and (b) the unitary-type Maxwell's demon. The dissipative information with an initial (c) classical correlation or (d) quantum correlation in terms of the evolution time as well as the noise parameters $\varepsilon_{c}$ or $\varepsilon_{q}$. {The classically or quantum correlated initial states are given by Eqs. (\ref{eq:rho_sr^c_noisy}) and (\ref{eq:rho_sr^q_noisy}), respectively.}}
\label{fig_example_2}
\end{figure}

{In Fig. \ref{fig_example_2}, we show the nonzero dissipative information based on the work extraction protocol given in Sec. \ref{sec:example_qubit_model}.} We can see that the dissipative information vanishes at $\varepsilon_{c}=1$, which gives rise to ${\Delta\mathcal S_{S|R}} = \Delta\mathcal S_{S}$ and ${\Sigma_{S|R}} = \Sigma_{S}$. The quantum part of the dissipative information vanishes at $\varepsilon_{q}=1$, while the dissipative information itself is nonzero because $\Sigma_{I}^{c}>0$. 
{Since we consider random implementations of the qubit Maxwell's demon, which cover all possibilities, we conclude that the nonzero dissipative information in the qubit Maxwell's demon during the work extraction protocol is universal.}

\section{Quantum fluctuation theorems of the conditional entropy production}

\label{sec:QFT}

Fluctuation theorems generalize the thermodynamic laws into a stochastic description, {where the inequality of the second law of thermodynamics generalizes to an equality valid for arbitrary nonequilibrium processes. The positivity of the averaged entropy production can be easily derived from the fluctuation theorem. Moreover, the fluctuation theorem also describes the statistics of entropy production \cite{merhavStatisticalPropertiesEntropy2010}. In quantum cases, the dynamics are replaced with stochastic trajectories, which are specified by the measurements performed at the beginning and the end of the dynamics } (the two-point measurement scheme) \cite{tasakiJarzynskiRelationsQuantum2000,espositoNonequilibriumFluctuationsFluctuation2009,campisiColloquiumQuantumFluctuation2011,funoQuantumFluctuationTheorems2018}. The back actions from the measurements are minimized if the measurements are performed at the eigenbasis of the system or the environment.

{Suppose that the initial state of the system and the reference have the decompositions
\begin{equation}
    \rho_S = \sum_n p_n|n\rangle_S\langle n|,\quad \rho_R = \sum_r p_r|r\rangle_R\langle r|,
\end{equation}
{with the corresponding eigenstate $|n\rangle_S$ ($|r\rangle_R$) and the probability of finding that eigenstate $p_n$ ($p_r$).} Since we have assumed a thermal state for the environment, we have
\begin{equation}
    \rho_E = \sum_m {\frac{e^{-\beta E_m}}{\mathcal Z_E}}|m\rangle_E\langle m|,
\end{equation}
with $E_m$ the energy of the state $|m\rangle$ {and $\mathcal Z_E$ the partition function of the environment.} The evolution $U_{SE}$ does not affect the reference state. Therefore, only the system has a different distribution after the evolution, given by
\begin{equation}
    \rho'_S = \sum_{n'}p_{n'}|n'\rangle_S\langle n'|.
\end{equation}
The statistics of the system are completely characterized by the distributions $p_n$ and $p_{n'}$.}

{The entropy of the state $\rho_S$ is given by $\mathcal S(\rho_S) = -\sum_n p_n\ln p_n$. We can then view the quantity $-\ln p_n$ as the unaveraged (stochastic) entropy. Following the same arguments, the stochastic version of the entropy production $\Sigma_S$ (\ref{def Sigma S}) is given by
\begin{equation}
    \sigma_S[\gamma_{SE}] =\Delta s_S +\beta q_E,
\end{equation}
with the stochastic entropy change $\Delta s_S = \ln p_n-\ln p_{n'}$ and the stochastic heat $q_E = E_{m'}-E_m$ \cite{deffnerNonequilibriumEntropyProduction2011}. Here, $\gamma_{SE} = \{n,m,n',m'\}$ is a set of stochastic variables.
}

{The stochastic trajectories give possible transitions from one initial microscopic state to one final microscopic state. The probability of one trajectory is then
\begin{equation}
\label{eq:p_f_se}
    P_{F}[\gamma_{SE}] = |{_{SE}\langle n',m'|}U_{SE}|n,m\rangle_{SE}|^2 p_n p_m,
\end{equation}
with {$p_m = e^{-\beta E_m}/\mathcal Z_E$}. The subscript F indicates that the trajectory is given by the forward evolution $U_{SE}$. The initial and final states are labeled $n,m$ and $n',m'$ respectively. Averaging the stochastic entropy production over all possible evolution trajectories gives the averaged quantity $\Sigma_S$ (\ref{def Sigma S}), i.e.,
\begin{equation}
\label{eq:sigma_s_average}
    \langle \sigma_S\rangle_{\gamma_{SE}} = \sum_{\gamma_{SE}}P_F[\gamma_{SE}] \sigma_S[\gamma_{SE}] = \Sigma_S.
\end{equation}
Therefore, we can also understand $P_F[\gamma_{SE}]$ as the probabilistic distribution of the stochastic entropy production $\sigma_S[\gamma_{SE}]$.
}

{The positivity of the entropy production $\langle \sigma_S\rangle_{\gamma_{SE}}\geq0$ only reflects the average of the stochastic entropy production. The fluctuation theorem captures the higher-order statistics of the stochastic entropy production with a concise equality \cite{landiIrreversibleEntropyProduction2021}
\begin{equation}
    \langle e^{-\sigma_S} \rangle_{\gamma_{SE}} = 1,
\end{equation}
which is called the integral fluctuation theorem. For the detailed version, see Sec. \ref{sec:experiments}. Applying Jensen's inequality $\langle e^{x}\rangle\geq e^{\langle x\rangle}$ to the above equality, we can get the averaged inequality $\langle \sigma_S\rangle_{\gamma_{SE}}\geq0$. The fluctuation theorem puts constraints on the higher-order statistics of the entropy production \cite{merhavStatisticalPropertiesEntropy2010}.}

{
Based on the definition of the conditional entropy production in Eq. (\ref{def SM}), we consider the stochastic conditional entropy production as 
\begin{equation}
\label{eq:def_sigma_s|r}
    \sigma_{S|R}[\gamma_{RSE}] = \Delta s_{S|R}+\beta q_{E},
\end{equation}
with the stochastic conditional entropy change $\Delta s_{S|R} = \ln(p_{n|r})-\ln(p_{n'|r})$. Here the conditional probability $p_{n|r}$ ($p_{n'|r}$) is given by $p_{n|r} = p_{n,r}/p_r$ ($p_{n'|r} = p_{n',r}/p_r$) with the joint distribution $p_{n,r} = {_{SR}\langle n,r|}\rho_{SR}|n,r\rangle_{SR}$ ($p_{n',r} = {_{SR}\langle n',r|}\rho'_{SR}|n',r\rangle_{SR}$). The stochastic variables are $\gamma_{RSE} = \{n,m,r,n',m'\}$. Since the local measurements commute with each other, the marginalization rules are guaranteed, i.e., $p_n = \sum_r p_{n,r}$ and $p_r = \sum_n p_{n,r}$. Note that the distribution of the reference $p_r$ does not change during the evolution $U_{SE}$. 
}

{
To include the reference, we extend the trajectory as
\begin{equation}
\label{eq:P_F_RSE}
    P_{F}[\gamma_{RSE}] = |{_{SE}\langle n',m'|}U_{SE}|n,m\rangle_{SE}|^2 p_{n,r} p_m,
\end{equation}
with the joint distribution $p_{n,r}$. The trajectory $P_{F}[\gamma_{RSE}]$ returns to $P_{F}[\gamma_{SE}]$ (\ref{eq:p_f_se}) when marginalizing over the variable $r$.} The measurements on the system and the {reference} commute, which guarantees the marginalization rules of the trajectory probability.

{
Since the conditional probability $p_{n|r}$ or $p_{n'|r}$ is obtained by measuring the system and the reference separately on their eigenstates, the quantum correlation is wiped out. Taking the average of the stochastic conditional entropy production $\sigma_{S|R}[\gamma_{RSE}]$ over the trajectory $P_{F}[\gamma_{RSE}]$ gives the averaged conditional entropy, which has a mismatch compared with $\Sigma_{S|R}$ (\ref{def SM}). Specifically, we have
\begin{align}
\label{eq:sigma_sr_average}
    \langle \sigma_{S|R}\rangle_{\gamma_{RSE}} = & \sum_{\gamma_{RSE}}P_F[\gamma_{RSE}] \sigma_{S|R}[\gamma_{RSE}] \nonumber \\
    = & \Sigma_{S|R}+\Delta J.
\end{align}
Here $\Delta J = J(\rho'_{SR})-J(\rho_{SR})$ with 
\begin{equation}
    J(\rho_{SR}) = \mathcal S(\tilde \rho_{SR}) - \mathcal S(\rho_{SR}).
\end{equation}
The density matrix $\tilde \rho_{SR}$ gives rise to dephasing because of the measurements, i.e., 
\begin{equation}
    \tilde \rho_{SR} =\sum_{n,r}|n,r\rangle_{SR}\langle n,r| \rho_{SR} |n,r\rangle_{SR}\langle n,r|,
\end{equation}
and this is also the case for the final state $\tilde\rho'_{SR}$. The entropy $J(\rho_{SR})$ is a coherence measure at the local basis $|n,r\rangle_{SR}$ \cite{baumgratzQuantifyingCoherence2014}. We have $\Delta J\leq 0$, since the final state has less coherence because of the coupling to the environment. When the initial state $\rho_{SR}$ is a classical correlated state, such as in Eq. (\ref{eq:rho_sr^c}), there are no back actions from the measurements, so $\Delta J=0$.
}

{Following the definition of the dissipative information in Eq. (\ref{eq SM S I}), the mismatch between $\sigma_{S|R}$ and $\sigma_{S}$ can be defined as the stochastic dissipative information, i.e.,
\begin{equation}
\label{eq:def_sigma_I}
    \sigma_I = \sigma_{S|R} - \sigma_{S}.
\end{equation}
Here $\sigma_I$ is also the stochastic mutual information change between the system and the reference \cite{parrondoThermodynamicsInformation2015}. Given the averaged relations in Eqs. (\ref{eq:sigma_s_average}) and (\ref{eq:sigma_sr_average}), we can easily see that
\begin{equation}
\label{eq:sigma_I_average}
    \langle \sigma_I\rangle_{\gamma_{RSE}} = \Sigma_I +\Delta J.
\end{equation}
Here $\langle \sigma_I\rangle_{\gamma_{RSE}}$ describes the dissipation of the classical correlation. 
}

{
We now show that the quantum fluctuation theorem based on the two-point measurement scheme can also be extended to conditional entropy production. Specifically, we have
}
\begin{theo}
\label{theorem 5}
\normalfont$\langle e^{-{\sigma_{S|R}}}\rangle_{{\gamma_{RSE}}\backslash{\gamma_{R}}} = \langle e^{-{\sigma_{S|R}}}\rangle_{{\gamma_{RSE}}} = 1$. The average over the variables ${\gamma_{RSE}}\backslash{\gamma_{R}}$ means taking the average over the trajectories conditioned on the state of the {reference}, i.e., 
\begin{equation}
    \langle \cdot\rangle_{{\gamma_{RSE}}\backslash{\gamma_{R}}} = \sum_{{\gamma_{RSE}}\backslash{\gamma_{R}}} \frac{P_{F}[{\gamma_{RSE}}]}{p_r}(\cdot).
\end{equation}
Here ${\gamma_{RSE}}\backslash{\gamma_{R}} = \{n,{m},n',{m'}\}$. In addition, we have $\langle\langle\cdot\rangle_{{\gamma_{RSE}}\backslash{\gamma_{R}}}\rangle_{{\gamma_{R}}} = \langle\cdot\rangle_{{\gamma_{RSE}}}$. 
\end{theo}
\begin{proof}
{The integral fluctuation theorem can be viewed as the normalization of a certain ``backward'' trajectory. It can be understood as a retrodiction process \cite{awFluctuationTheoremsRetrodiction2021,buscemiFluctuationTheoremsBayesian2021}. Consider the backward trajectory corresponding to the forward trajectory (\ref{eq:P_F_RSE}):
\begin{equation}
    P_{B}[\gamma_{RSE}] =  |{_{SE}\langle n,m|}U^\dag_{SE}|n',m'\rangle_{SE}|^2 p_{n'r}p_{m'},
\end{equation}
with $p_{n'r} = {_{SR}\langle n',r|}\rho'_{SR}|n',r\rangle_{SR}$ and $p_{m'} = {_E\langle m'|}\rho_E|m'\rangle_E$. Note that the initial state of the backward trajectory is not the final state of the forward evolution. If we marginalize over the variable $r$ of $P_{B}[\gamma_{RSE}]$, we get the backward trajectory corresponding to the forward trajectory $P_F[\gamma_{SE}]$ (\ref{eq:p_f_se}). }

First, we can verify the following normalization conditions
\begin{subequations}
\begin{align}
    &\sum_{{\gamma_{RSE}}} P_{F}[{\gamma_{RSE}}] = \sum_{{\gamma_{RSE}}}P_{B}[{\gamma_{RSE}}]= 1; \\
    &\sum_{{\gamma_{SE}}} P_{F}[{\gamma_{SE}}] = \sum_{{\gamma_{SE}}} P_{B}[{\gamma_{SE}}] = 1.
\end{align}
\end{subequations}
Moreover, the trajectory $P_{F}[{\gamma_{RSE}}]$ conditioned on the probability of the {reference} also gives the normalization
\begin{multline}
    \sum_{{\gamma_{RSE}}\backslash{\gamma_{R}}} \frac{P_{F}[{\gamma_{RSE}}]}{{p_r}} \\
    = \sum_{\gamma_{SE}}\frac{{|{_{SE}\langle n,m|}U^\dag_{SE}|n',m'\rangle_{SE}|^2} {p_{n,r}p_m}}{{p_r}} \\  = \sum_n\frac{{p_{n,r}}}{{p_r}} = 1.
\end{multline}
Similar normalization can be applied to the backward trajectory. 

{Next, we can rewrite the conditional stochastic entropy production $\sigma_{S|R}$ as}
\begin{align}
    {\sigma_{S|R}[\gamma_{RSE}]} = & \ln\left(\frac{{p_{n,r}p_m}}{{p_r}}\right) - \ln\left(\frac{{p_{n',r}p_{m'}}}{{p_r}}\right) \nonumber \\
    = & \ln\left( \frac{P_{F}[{\gamma_{RSE}}]}{p_r}\right) - \ln\left(\frac{P_{B}[{\gamma_{RSE}}]}{p_r}\right).
\end{align}
We then have the integral fluctuation relation
\begin{multline}
    \langle e^{-{\sigma_{S|R}}}\rangle_{{\gamma_{RSE}}\backslash{\gamma_{R}}} =  \sum_{{\gamma_{RSE}}\backslash{\gamma_{R}}}\frac{P_{F}[{\gamma_{RSE}}]}{p_r} \frac{P_{B}[{\gamma_{RSE}}]}{P_{F}[{\gamma_{RSE}}]}
    \\
    =  \sum_{{\gamma_{RSE}}\backslash{\gamma_{R}}}\frac{P_{B}[{\gamma_{RSE}}]}{p_r}
    =  1.
\end{multline}
Furthermore, we have
\begin{equation}
    \langle e^{-{\sigma_{S|R}}}\rangle_{{\gamma_{RSE}}} = \langle\langle e^{-{\sigma_{S|R}}}\rangle_{{\gamma_{RSE}}\backslash{\gamma_{R}}}\rangle_{{\gamma_{R}}} = \sum_r p_r = 1.
\end{equation}
\end{proof}

More interestingly, the stochastic dissipative information itself $\sigma_I$, given by Eq. (\ref{eq:def_sigma_I}), also follows the fluctuation theorem.
\begin{theo}
\label{theorem 6}
\normalfont$\langle e^{-\sigma_{I}}\rangle_{{\gamma_{RSE}}\backslash{\gamma_{SE}}} =\langle e^{-\sigma_{I}}\rangle_{{\gamma_{RSE}}} = 1$.
The quantum fluctuation theorem of dissipative information is given by the conditional trajectories, i.e., 
\begin{equation}
    \langle \cdot\rangle_{{\gamma_{RSE}}\backslash{\gamma_{SE}}} = \sum_{{\gamma_{RSE}}\backslash{\gamma_{SE}}} P_{F}[{\gamma_{RSE}}]/P_{F}[{\gamma_{SE}}](\cdot).
\end{equation}
\end{theo}
\begin{proof}
The stochastic dissipative information $\sigma_I$ can be rewritten as
\begin{equation}
    \sigma_{I}[\gamma_{RSE}] = \ln\left(\frac{P_{F}[{\gamma_{RSE}}]}{P_{F}[{\gamma_{SE}}]}\right) - \ln\left(\frac{P_{B}[{\gamma_{RSE}}]}{P_{B}[{\gamma_{SE}}]}\right).
\end{equation}
We then have
\begin{multline}
    \langle e^{-\sigma_{I}}\rangle_{{\gamma_{RSE}}\backslash{\gamma_{SE}}} =  \sum_{{\gamma_{RSE}}\backslash{\gamma_{SE}}}\frac{P_{F}[{\gamma_{RSE}}]}{P_{F}[{\gamma_{SE}}]} \frac{P_{F}[{\gamma_{SE}}]P_{B}[{\gamma_{RSE}}]}{P_{F}[{\gamma_{RSE}}]P_{B}[{\gamma_{SE}}]} \\
    = \sum_{r}\frac{P_{B}[{\gamma_{RSE}}]}{P_{B}[{\gamma_{SE}}]} = \sum_r\frac{P_{n,r}}{P_n} = 1.
\end{multline}
Furthermore, we have
\begin{equation}
    \langle e^{-\sigma_{I}}\rangle_{{\gamma_{RSE}}} = \langle\langle e^{-\sigma_{I}}\rangle_{{\gamma_{RSE}}\backslash{\gamma_{SE}}}\rangle_{{\gamma_{SE}}} = \sum_{{\gamma_{SE}}}P_{F}[{\gamma_{SE}}] = 1.
\end{equation}
\end{proof}

{The fluctuation theorem of the dissipative information guarantees the positivity of $\langle \sigma_{I}\rangle_{\gamma_{RSE}}\geq 0$, which is equal to the classical part of $\Sigma_I$ as suggested by Eq. (\ref{eq:sigma_I_average}). When we apply the global two-point measurement to preserve the quantum correlation between the system and the reference, we can obtain the fluctuation theorem of the full quantum conditional entropy production $\Sigma_{S|R}$ (see Sec. \ref{sec:ft_global}). However, the fluctuation theorem of the full quantum dissipative information requires further treatment.} As pointed out in \cite{yungerhalpernJarzynskilikeEqualityOutoftimeordered2017,kwonFluctuationTheoremsQuantum2019,levyQuasiprobabilityDistributionHeat2020}, the fluctuation theorem about the nonclassical correlation should be described by the quasiprobability instead of the probability. The fluctuation theorem of the dissipative information given by the quasiprobability implies the strong subadditivity of the von Neumann entropy, which is beyond the scope of our current paper.





\section{\label{sec:ft_global} The fluctuation theorem of the conditional entropy production with the global two-point measurement}

{When we introduced the conditional entropy production in Sec. \ref{sec:conditional_ep},} both classical and quantum correlations between the system and the {reference} were included. {In Sec. \ref{sec:QFT}, we applied the two-point measurement and established the fluctuation theorem of the conditional entropy production and the dissipative information. However, local measurements were applied to the system and the reference. Therefore, their quantum correlations were wiped out. In this section, we show this issue can be circumvented by applying the global two-point measurement \cite{parkFluctuationTheoremArbitrary2017,micadeiQuantumFluctuationTheorems2020,parkInformationFluctuationTheorem2020}.}

{Suppose that the initial and final states have the decompositions
\begin{equation}
    \rho_{SR} = \sum_l p_l|l\rangle_{SR}\langle l|,\quad \rho'_{SR} = \sum_{l'} p_{l'}|l'\rangle_{SR}\langle l'|,
\end{equation}
with the entangled eigenstates $|l\rangle_{SR}$ and $|l'\rangle_{SR}$. We then define the stochastic conditional entropy production as
\begin{equation}
    \sigma^q_{S|R}[\gamma^q_{RSE}] = \Delta s^q_{S|R}+\beta q_E,
\end{equation}
with the stochastic conditional entropy change $\Delta s^q_{S|R} = -\ln(p_l/p_r)-(-\ln p_{l'}/p_r)$ and the stochastic heat $q_E = E_{m'}-E_m$. The stochastic variables are $\gamma^q_{RSE} = \{l,m,r,l',m'\}$.  Note that $\sigma^q_{S|R}$ and $\Delta s^q_{S|R}$ are different from those defined in Eq. (\ref{eq:def_sigma_s|r}). We use a superscript q to denote that the quantum correlation is included. Correspondingly, the stochastic dissipative information becomes
\begin{equation}
\label{eq:def_sigma^q_I}
    \sigma^q_I = \sigma^q_{S|R} - \sigma_{S},
\end{equation}
which characterizes the change in the quantum stochastic mutual information \cite{parkFluctuationTheoremArbitrary2017,micadeiQuantumFluctuationTheorems2020,parkInformationFluctuationTheorem2020}. 
}

{Corresponding to the stochastic entropy production $\sigma^q_{S|R}[\gamma^q_{RSE}]$, we modify the trajectory as}
{
\begin{equation}
    P_{F}[\gamma^{q}_{RSE}] = |{_{RSE}\langle l',m'|}U_{SE}|l,m\rangle_{RSE}|^2 p_l p_m.
\end{equation}}
Similarly, the backward trajectory given by ${U_{SE}^\dag}$ is
{
\begin{equation}
    P_{B}[\gamma^{q}_{RSE}] = |{_{RSE}\langle l,m|}U^\dag_{SE}|l',m'\rangle_{RSE}|^2 p_{l'} p_{m'},
\end{equation}
with $p_{l'} p_{m'} = {_{RSE}\langle l',m'|}\rho_{SR}'\otimes \rho_{E}|l',m'\rangle_{RSE}$. Since there are no back actions from the initial measurements, one can verify that the averaged stochastic entropy production is 
\begin{equation}
\label{eq:average_sigma_sm}
    \langle \sigma^q_{S|R}\rangle_{\gamma^{q}_{RSE}} =  \Sigma_{S|R},
\end{equation}
which is different from Eq. (\ref{eq:sigma_sr_average}).
}

Equipped with the stochastic conditional entropy production and the global trajectory, we have the integral quantum fluctuation theorem:
\begin{theo}
\label{theorem 4}
\normalfont$\langle {e^{-\sigma^{q}_{S|R}}}\rangle_{{\gamma^{q}_{RSE}}} = \langle e^{-(\sigma_{S}+\sigma^{q}_{I})}\rangle_{{\gamma^{q}_{RSE}}} = 1$.
The average notation means $\langle\cdot\rangle_{{\gamma^{q}_{RSE}}}  = \sum_{{\gamma_{RSE}^{q}}}P_{F}[{\gamma^{q}_{RSE}}](\cdot)$. 
\end{theo}
\begin{proof}
Both the forward and backward global trajectories are normalized
\begin{equation}
    \sum_{{\gamma^{q}_{RSE}}} P_{F}[{\gamma^{q}_{RSE}}] = \sum_{{\gamma^{q}_{RSE}}}P_{B}[{\gamma^{q}_{RSE}}]=1.
\end{equation}
The stochastic conditional entropy production $\sigma^{q}_{S|R}$ can be written using the forward and backward trajectories 
{
\begin{align}
    \sigma^q_{S|R}[\gamma^q_{RSE}] = & \ln\left(\frac{p_{l}p_m}{p_r}\right) - \ln\left(\frac{p_{l'}p_{m'}}{p_r}\right) \nonumber \\
    = & \ln\left( \frac{{P_{F}[\gamma^{q}_{RSE}}]}{p_r}\right) - \ln\left(\frac{{P_{B}[\gamma^{q}_{RSE}}]}{p_r}\right).
\end{align}}
We then have
\begin{align}
    \langle e^{-{\sigma^q_{S|R}}}\rangle_{{\gamma^{q}_{RSE}}} = & \langle e^{-(\sigma_{S}+{\sigma^q_{I}})}\rangle_{{\gamma^{q}_{RSE}}} \nonumber \\
    = &  \sum_{{\gamma^{q}_{RSE}}} P_{F}[{\gamma^{q}_{RSE}}] \frac{P_{B}[{\gamma^{q}_{RSE}}]}{P_{F}[{\gamma^{q}_{RSE}]}}\nonumber \\
    = & \sum_{{\gamma^{q}_{RSE}}}P_{B}[{\gamma^{q}_{RSE}}] \nonumber \\
    = & 1.
\end{align}
\end{proof}
Note that the quantum correlation between the system and the {reference} is included in the above quantum fluctuation theorem. Applying Jensen's inequality gives $\langle {\sigma^q_{S|R}}\rangle_{{\gamma^{q}_{RSE}}}\geq 0$, which also proves the positivity of the conditional entropy production $\Sigma_{S|R}$. {Unlike the stochastic dissipative information $\sigma_I$ (\ref{eq:def_sigma_I}) given by the local two-point measurement scheme, the stochastic dissipative information $\sigma_I^q$ including the quantum correlation does not follow a fluctuation theorem like $\sigma_I$. The issue arises from the classical description of the trajectories, which is incapable of describing the dynamics of quantum information. A possible solution is to define the quasiprobability trajectory, but we leave this for future study.   }

\section{Experimental verifications of the new quantum fluctuation theorem on IBM quantum computers}

\label{sec:experiments}

Quantum fluctuation theorems have previously been verified on different platforms via the interferometric method \cite{dornerExtractingQuantumWork2013,batalhaoExperimentalReconstructionWork2014,cerisolaUsingQuantumWork2017} or the two-point projective measurements method \cite{anExperimentalTestQuantum2015,masuyamaInformationtoworkConversionMaxwell2018}. Recent studies have demonstrated the quantum fluctuation theorem on quantum computers \cite{solfanelliExperimentalVerificationFluctuation2021}. State-of-the-art quantum computers provide a powerful platform with which to verify quantum fluctuation theorems based on the dynamics of several qubits. In this section, we study the experimental verification of Theorems \ref{theorem 5} and \ref{theorem 6} on IBM quantum computers \cite{IBM}.

We design a three-qubit example to verify the quantum fluctuation theorem of dissipative information. The system S, {the reference R}, and the environment {E} are all one qubit. We set the initial state of {SR} as ${\rho_{SR}^{c}(\varepsilon_{c})}$, as shown in Eq. (\ref{eq:rho_sr^c_noisy}). {Therefore, there are no back actions from the two-point measurements.} Both qubits S and {R} have the local eigenstates $|0\rangle$ and $|1\rangle$. The statistics of the state ${\rho_{SR}^{c}(\varepsilon_{c})}$ (based on measurements on the computational basis) are equivalent to an entangled pure state with the same diagonal density matrix. We design the following evolution to prepare such a pure state:
\begin{equation}
\Qcircuit @C=0.7em @R=1.5em {
 & \multigate{1}{{U_{SR}}} \qw & \qw  &&&&& \gate{R_y(\theta_1)} & \ctrl{1} & \gate{R_y(\theta_2)} & \qw \\
 & \ghost{{U_{SR}}} \qw & \qw &\raisebox{1.1cm}{~~~=} &&&& \qw & \targ & \gate{R_y(\theta_2)} & \qw\\
}
\end{equation}
with a single-qubit rotation $y-$axis gate $R_y(\theta) = e^{-i\theta\sigma^y/2}$ and a two-qubit CNOT gate \cite{nielsenQuantumComputationQuantum2010}. The angles are determined by
\begin{equation}
\begin{dcases}
p = \frac 1 2 \left(\cos^2\left(\frac{\theta_1+\theta_2}{2}\right)+\cos^2\left(\frac{\theta_1-\theta_2}{2}\right)\right);\\ 
\varepsilon_{c}p(1-p) = \frac 1 2 \sin^2\theta_2\sin^2\left(\frac{\theta_1}{2}-\frac{\pi}{4}\right).
\end{dcases}
\end{equation}


The {environmental qubit E} is thermal, and can be prepared by evolution on $|00\rangle_{EV}$:
\begin{equation}
\Qcircuit @C=0.7em @R=1.5em {
 & \multigate{1}{{U_{EV}}} \qw & \qw  &&&&& \gate{R_y(\theta_3)} & \ctrl{1} & \qw \\
 & \ghost{{U_{EV}}} \qw & \qw &\raisebox{1.1cm}{~~~=} &&&& \qw & \targ & \qw\\
}
\end{equation}
where the angle $\theta_3 = 2\arctan(e^{\beta})$. Here the qubit V is an ancillary qubit.

The evolution ${U_{{SE}}}\in SO(4)$, given by Eq. (\ref{eq:XY_interaction}), can be optimally realized by two CNOT gates plus the single-qubit gates \cite{vatanOptimalQuantumCircuits2004}:
\begin{equation}
\Qcircuit @C=0.7em @R=1.5em {
 & \multigate{1}{{U_{SE}}} \qw & \qw  &&&&& \gate{S} & \gate{H} & \ctrl{1} & \gate{R_y(2g)} & \ctrl{1} & \gate{H} & \gate{S^\dag} & \qw \\
 & \ghost{{U_{SE}}} \qw & \qw &\raisebox{1.1cm}{~~~=} &&&& \gate{S} & \qw &\targ & \gate{R_y(2g)} & \targ & \gate{S^\dag} & \qw & \qw \\
}
\end{equation}
with the single-qubit phase gate $S = \text{diag}\{0,i\}$ and the Hadamard gate $H = (\sigma^x+\sigma^z)/\sqrt 2$.


\begin{figure}[t]
$$
\Qcircuit @C=0.5em @R=1.5em {
|0\rangle_{R} &&& \multigate{1}{{U_{SR}}} & \meter \\
|0\rangle_{S} &&& \ghost{{U_{SR}}} & \meter & \multigate{1}{{U_{SE}}} & \meter & \qw & \qw & \qw & \qw & \qw & \qw & \multigate{1}{{U^\dag_{SE}}} & \meter \\
|0\rangle_{E} &&& \multigate{1}{{U_{EV}}} &  \meter & \ghost{{U_{SE}}} & \meter && |0\rangle_{E} &&& \multigate{1}{{U_{EV}}} & \meter & \ghost{{U^\dag_{SE}}} & \meter \\ 
|0\rangle_{V} &&& \ghost{{U_{EV}}} & \qw &&&&  |0\rangle_{V} &&& \ghost{{U_{EV}}} 
}
$$
\caption{Quantum circuits designed based on the two-point measurement scheme. The evolution ${U_{SR}}$ prepares the correlated qubits (system and {reference}). The evolution ${U_{EV}}$ prepares the thermal environmental qubit {E}. The evolution ${U_{SE}}$ is set as the XY interaction between the two qubits given by Eq. (\ref{eq:XY_interaction}).}
\label{fig_ft_qc}
\end{figure}

\begin{figure}[t]
\centering
\includegraphics[width=0.23\textwidth]{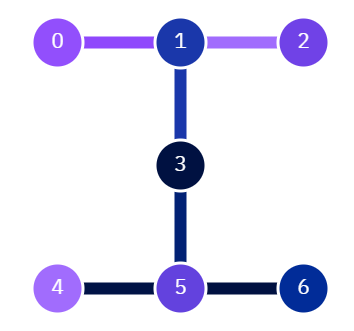}
\caption{The qubits layout of the IBM quantum processor {\fontfamily{qcr}\selectfont ibm\_lagos}. The colors of the connected lines represent the two-qubit gate errors. A darker color means a lower error rate.}
\label{fig_lagos}
\end{figure}

According to the two-point measurement scheme, we perform local eigenstate measurements on the system, reference, and environment. After the initial measurement, the system S and the environment {E} undergo the evolution ${U_{{S}{E}}}$ (realized by the single- and two-qubit gates). Lastly, the final projective measurements are applied. Note that the state of the {reference} does not change, and therefore, only a one-point projective measurement is applied on the {reference qubit R}. We can continuously implement the backward process on the final state, given by the evolution ${U^\dag_{{S}{E}}}$ (see Fig. \ref{fig_ft_qc}). IBM quantum computers allow intermediate measurements in the circuits. One can also use separate circuits, which separately measure the initial and final state statistics, as well as the dynamics of the evolution \cite{herreraEasyAccessEnergy2021}. We follow the latter method to avoid accumulated errors. Note that the environmental qubit has the same initial state as that in the forward process. We apply the designed circuits on the IBM quantum processor {\fontfamily{qcr}\selectfont ibm\_lagos}. See Fig. \ref{fig_lagos} for the qubits layout of this processor. We map the qubits {R}, S, {E}, and V to the qubits 1, 3, 5, and 6. The data below were collected (via the dedicated mode) on Aug. 12 2021 from 9:45 pm to 10:00 pm (GMT-4). The metrics of the processor {\fontfamily{qcr}\selectfont ibm\_lagos} during the above time were as follows: average CNOT errors: {$0.6441\%$}; average readout errors: {$1.031\%$}; average T1 time: 122.88 {$\mu s$}; average T2 time: 74.81 {$\mu s$}.

\begin{figure*}[t]
\centering
\includegraphics[width=0.9\textwidth]{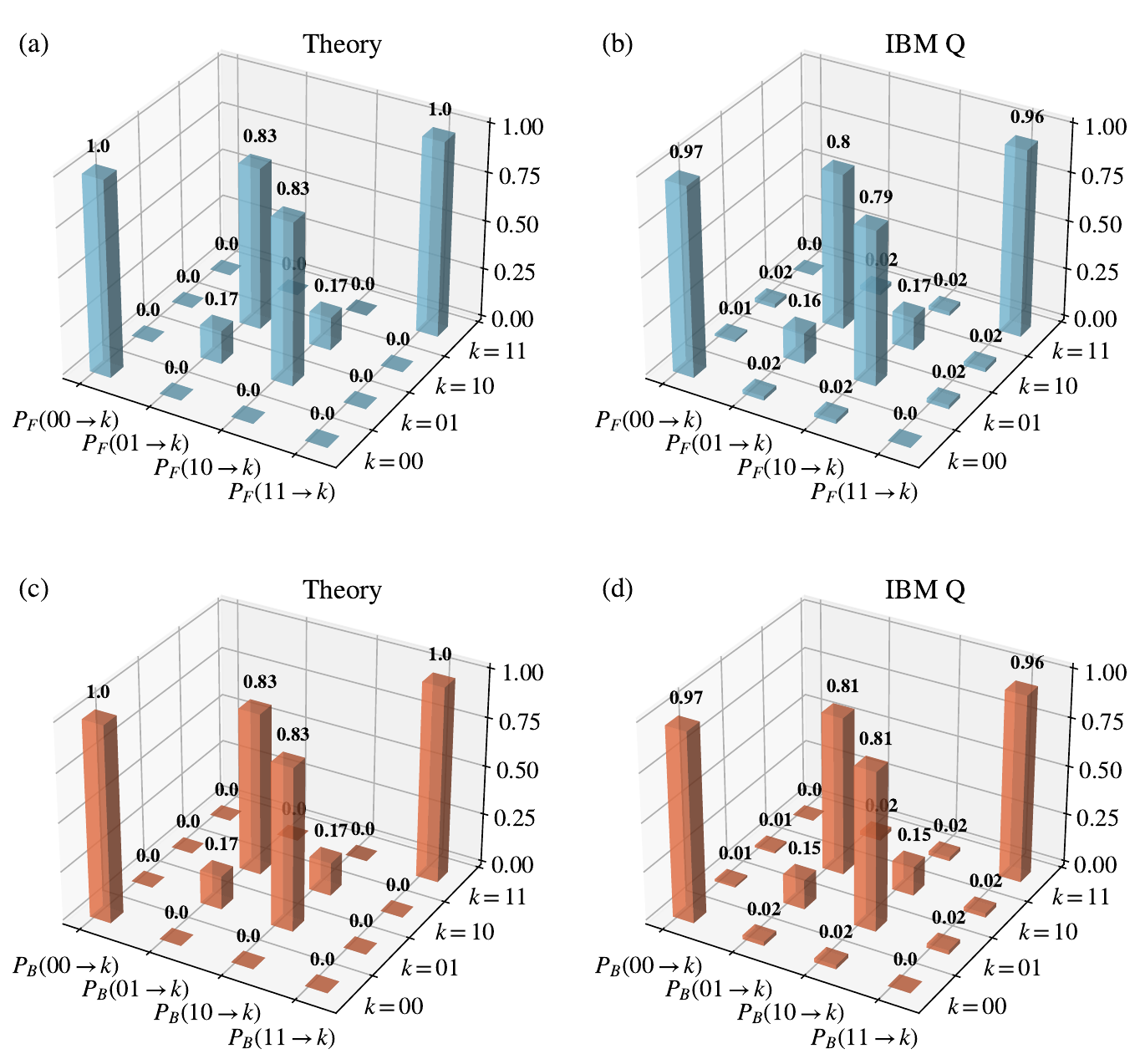}
\caption{The transition probabilities of the forward and backward processes. The evolution coupling constant is set as $g=1$. (a) and (c) show the theoretical values for the forward and backward processes. (b) and (d) show the experimental values measured on the {\fontfamily{qcr}\selectfont ibm\_lagos} processor.}
\label{fig_trans}
\end{figure*}

\begin{figure*}
\includegraphics[width=1\textwidth]{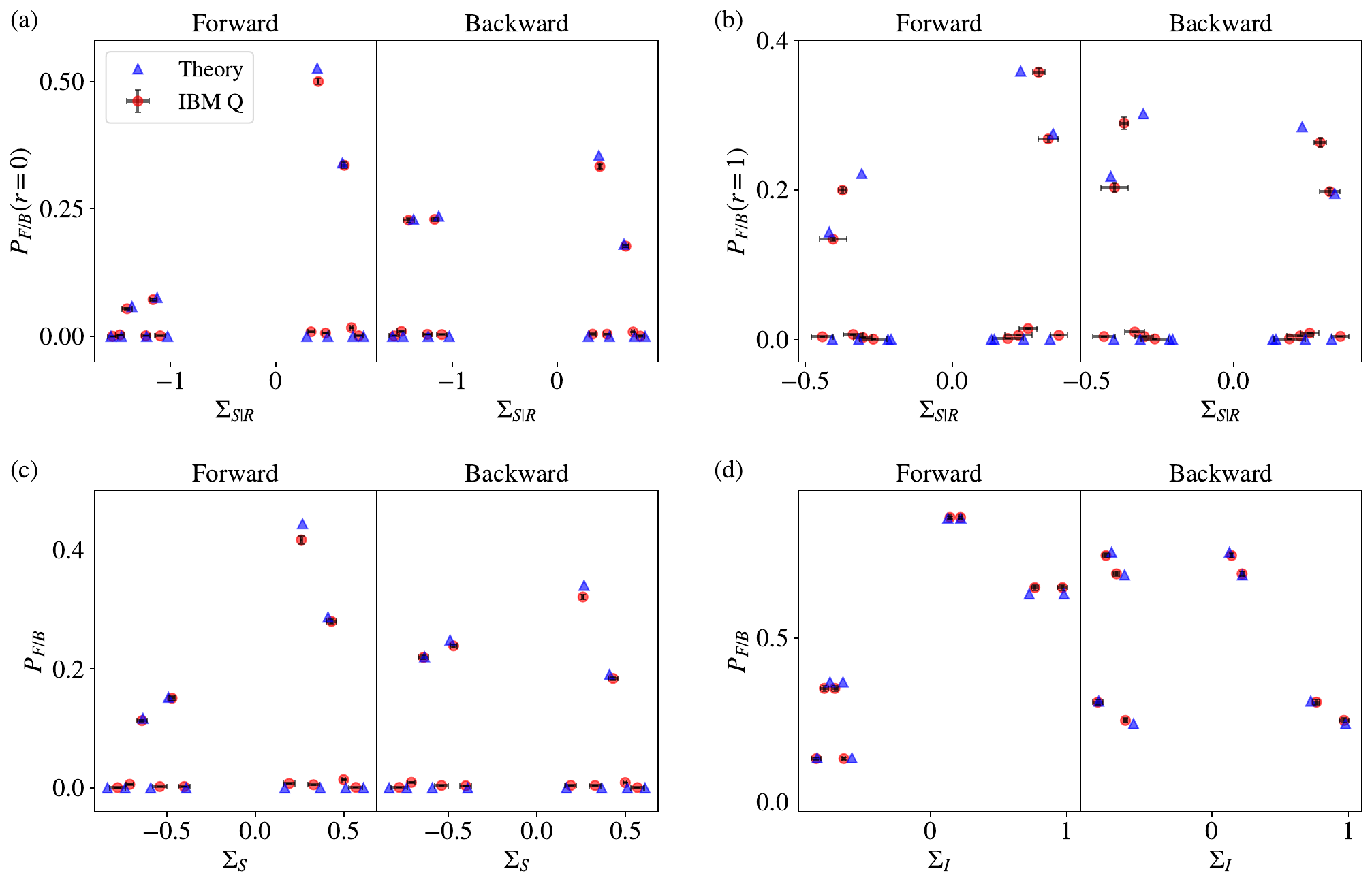}
\caption{The probabilistic distributions of (a),(b) the conditional entropy production ${\sigma_{S|R}}$, (c) the unconditional entropy production $\sigma_{S}$, and (d) the dissipative information $\sigma_{I}$. The parameters are set as $\varepsilon_{c} = 0.5$, $\beta_{S} = 10\beta$, and $g=1$. The statistical variance is based on $5\times8192$ shots of circuits on the IBM quantum computer {\fontfamily{qcr}\selectfont ibm\_lagos}. }
\label{fig_ep_p}
\end{figure*}

\begin{figure*}
\includegraphics[width=1\textwidth]{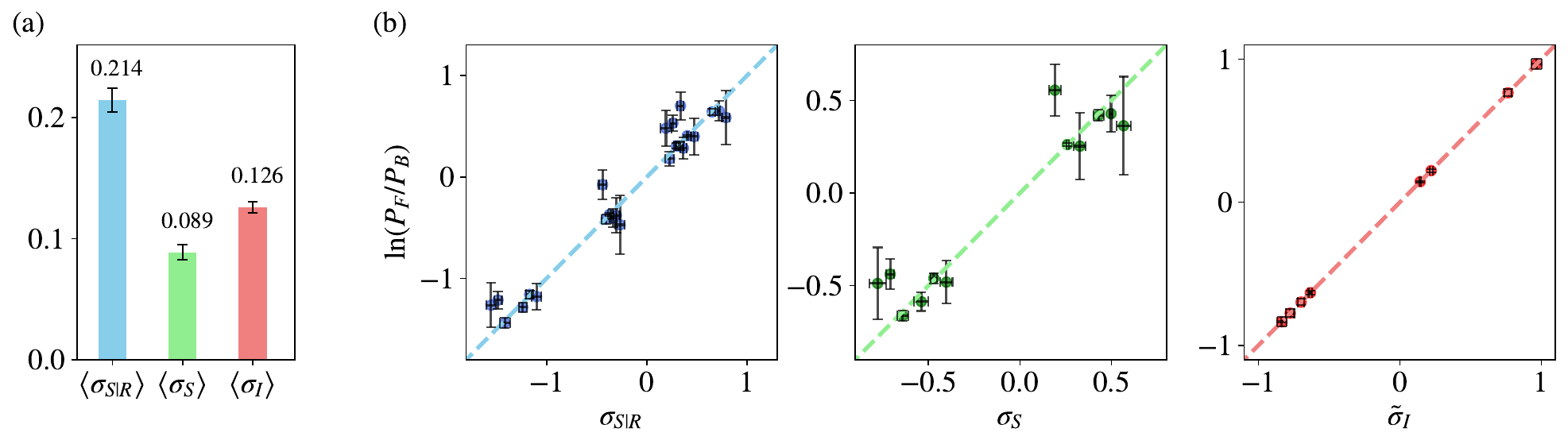}
\caption{(a) The averaged values of the conditional and unconditional entropy production as well as the dissipative information. (b) Verification of the detailed fluctuation theorems in terms of {$\sigma_{S|R}$}, $\sigma_{S}$, and $\sigma_{I}$. The parameters are set as $\varepsilon_{c} = 0.5$, $\beta_{S} = 10\beta$, and $g=1$. The statistical variance is based on $5\times8192$ shots of circuits on the IBM quantum computer {\fontfamily{qcr}\selectfont ibm\_lagos}.}
\label{fig_ft_ibm}
\end{figure*}

To get the trajectories $P_F[\gamma_{RSE}]$ and $P_B[\gamma_{RSE}]$ of the evolution, we first measure the transition probabilities, defined as 
\begin{equation}
P_{F}(j\rightarrow k) = |\langle k|{U_{SE}}|j\rangle|^2,\quad P_{B}(j\rightarrow k) = |\langle k|{U^\dag_{SE}}|j\rangle|^2,
\end{equation}
with $|j\rangle,|k\rangle\in\{|00\rangle,|01\rangle,|10\rangle,|11\rangle\}$. The theoretical and measured transition probabilities are shown in Fig. \ref{fig_trans}. The probabilities are estimated by the statistics of $5\times8192$ shots of the circuits. Here 8192 is the maximal number of shots allowed on the IBM Q platform. We repeat the running process five times. 

Besides the transition probabilities, we measure the statistics of the initial and the final states (two-point measurements), which give the stochastic entropy production parameters ${\sigma_{S|R}}$, $\sigma_{S}$, and $\sigma_{I}$ (with the relation ${\sigma_{S|R}} = \sigma_{S}+\sigma_{I}$). Combining the transition probabilities of the forward and the backward processes with the statistics of the initial and final states, we get the forward and backward trajectories. 

Instead of experimentally verifying the integral quantum fluctuation theorems presented in Theorems \ref{theorem 5} and \ref{theorem 6}, we verify their corresponding detailed fluctuation theorems, which are more experiment-friendly. Note that the detailed fluctuation theorems automatically give rise to the integral fluctuation theorem (after the summation of all trajectories). Consider the probabilistic distribution of the unconditional entropy production $\sigma_{S}$
\begin{equation}
    P_{F}(\Sigma_{S}) = \sum_{{\gamma_{SE}}} P_{F}[{\gamma_{SE}}]\delta(\sigma_{S},\Sigma_{S}),
\end{equation}
where $\delta$ is the Kronecker delta. The probabilistic distribution of the backward process can be defined similarly, where the positive entropy production in the backward process is interpreted as the negative entropy production in the forward process. We then have the detailed quantum fluctuation theorem
\begin{equation}
\label{eq:dft_sigma_s}
    \Sigma_{S} = \ln\frac{P_{F}(\Sigma_{S})}{P_{B}(-\Sigma_{S})},
\end{equation}
which naturally gives $\langle e^{-\sigma_{S}}\rangle_{{\gamma_{SE}}} = 1$. 

The probabilistic distribution of the conditional entropy production is based on the conditional trajectories (conditioned on the microstate of the {reference} denoted by $r$) and is given by
\begin{equation}
    P_{F}({\Sigma_{S|R}},r) = \sum_{{\gamma_{SE}}}\frac{P_{F}[{\gamma_{RSE}}]}{p_r} \delta({\sigma_{S|R}},{\Sigma_{S|R}}).
\end{equation}
We then have the detailed quantum fluctuation theorem
\begin{equation}
\label{eq:dft_sigma_sm}
    {\Sigma_{S|R}} = \ln\frac{P_{F}({\Sigma_{S|R}},r)}{P_{B}(-{\Sigma_{S|R}},r)}.
\end{equation}
When the {reference} is a qubit, we have two probabilistic distributions of the conditional entropy production in terms of $r=0$ or $r=1$ (see Fig. \ref{fig_ep_p}).

The stochastic dissipative information has a distribution based on the conditional trajectories (conditioned on the trajectories of {SE}) given by
\begin{equation}
    P_{F}(\Sigma_{I},{\gamma_{SE}}) = \sum_{r}\frac{P_{F}[{\gamma_{RSE}}]}{P_{F}[{\gamma_{SE}}]} \delta(\sigma_{I},\Sigma_{I}).
\end{equation}
The detailed quantum fluctuation theorem for the stochastic dissipative information is then given by
\begin{equation}
\label{eq:dft_sigma_i}
    \Sigma_{I} = \ln\frac{P_{F}(\sigma_{I},{\gamma_{SE}})}{P_{B}(-\sigma_{I},{\gamma_{SE}})},
\end{equation}
which gives the integral fluctuation theorem $\langle e^{-\sigma_{I}}\rangle_{{\gamma_{RSE}}\backslash{\gamma_{SE}}} =1$. 

Based on the statistical distributions in Fig. \ref{fig_ep_p}, we plot the average values of the three types of entropy production in Fig. \ref{fig_ft_ibm}(a), which verify the relation 
\begin{equation}
    \langle {\sigma_{S|R}}\rangle_{\gamma_{RSE}} = \langle\sigma_{S}\rangle_{\gamma_{RSE}}+\langle\sigma_{I}\rangle_{\gamma_{RSE}}.
\end{equation}
Fig. \ref{fig_ft_ibm}(b) shows the detailed fluctuation theorems in terms of ${\sigma_{S|R}}$, $\sigma_{S}$, and $\sigma_{I}$ (based on the different statistics of the forward and backward processes). The experimental values unambiguously show that each of them follows the quantum fluctuation theorem. 

{In the above experiments, the evolution $U_{SE}$ has zero transition probabilities between several states (the forbidden transitions; see Fig. \ref{fig_trans}). However, the measured transition probabilities are all nonzero due to the imperfect operations of quantum computers. The theoretically forbidden trajectories fluctuate much more than the theoretically well-defined ones, which leads to deviations of $\sigma_{S|R}$ and $\sigma_{S}$, as seen in Fig. \ref{fig_ft_ibm}(b) (with relatively large error bars). As proposed in \cite{solfanelliExperimentalVerificationFluctuation2021}, it would be interesting to explore this further in the future on benchmarking quantum computers via quantum fluctuation theorems. }

{On the other hand, the error bars of the fluctuation relation in terms of $\sigma_{I}$ in Fig. \ref{fig_ft_ibm}(b) vanish, since the transition probabilities cancel each other out in the conditional trajectories $P_{F}[\gamma_{RSE}]/P_{F}[\gamma_{SE}]$ and $P_{B}[\gamma_{RSE}]/P_{B}[\gamma_{SE}]$. More specifically, the ratio $\ln(P_{F}/P_{B})$ in terms of $\sigma_{I}$ is equal to the definition of $\sigma_{I}$ directly. Note that we apply the same initial and final distributions to evaluate the stochastic dissipative information and the corresponding forward and backward trajectories.}

\section{Conclusion}

Information loss is subjective, which motivates the study of entropy production from different perspectives. In this study, we have investigated a quantum system that is initially correlated with a reference and interacts with a thermal environment. The irreversibility of the dynamics is different depending on whether the viewpoint is that of the system or the reference. The system-environment interaction has a positive conditional entropy production with respect to the reference (conditioned on the state of the reference), which is always larger than the unconditional entropy production with respect to the system. We show that when the system reaches thermal equilibrium, namely the system-environment interaction has a zero unconditional entropy production (with respect to the system), it can still have a nonzero conditional entropy production with respect to the reference. Therefore, the conditional entropy production delineates the thermal nonequilibrium and the informational nonequilibrium dynamics.



Compared to the unconditional entropy production with respect to the system, conditional entropy production with respect to the reference has an additional contribution, which is called the dissipative information in our study. The dissipative information quantifies the thermodynamic cost associated with maintaining the system-reference correlation when the system is coupled to the environment. We have shown that the dissipative information is equal to the mutual information change between the system and the reference during the system-environment dynamics. It is also equivalent to the amount of final-state conditional mutual information between the reference and the environment. The dissipative information then provides a thermodynamic interpretation of the conditional mutual information \cite{fawziQuantumConditionalMutual2015,brandaoQuantumConditionalMutual2015}. Positive dissipative information also suggests wasted correlation work \cite{funoThermodynamicWorkGain2013,beraGeneralizedLawsThermodynamics2017,manzanoOptimalWorkExtraction2018}. We have demonstrated universal nonzero dissipative information in the qubit Maxwell's demon model, which cannot be revealed in the unconditional entropy production.

In addition, we have established quantum fluctuation theorems of the conditional entropy production based on the two-point measurement scheme \cite{tasakiJarzynskiRelationsQuantum2000,espositoNonequilibriumFluctuationsFluctuation2009,campisiColloquiumQuantumFluctuation2011,funoQuantumFluctuationTheorems2018}. {Moreover, we have shown that the dissipative information itself also follows a fluctuation theorem beyond the fluctuation theorem of the thermodynamic quantities. This suggests that the fluctuation theorem is a universal tool with which to study the dynamics and statistics of quantum information.} Based on the dynamics of three qubits, we have verified the new quantum fluctuation theorem for the conditional entropy production and the dissipative information experimentally on an IBM quantum computer. This study on conditional entropy production and dissipative information clarifies informational nonequilibrium \cite{parrondoThermodynamicsInformation2015}. It enriches our understanding of the nonequilibrium thermodynamics of quantum information processing tasks.



\quad

\begin{acknowledgments}

We acknowledge access to advanced services provided by the IBM Quantum Researchers Program. In this paper, we used {\fontfamily{qcr}\selectfont ibm\_lagos}, which is one of the IBM Quantum Canary Processors. The authors thank Wei Wu, Wufu Shi, and Hong Wang for helpful discussions.
	
\end{acknowledgments}


\begin{thebibliography}{65}%
\makeatletter
\providecommand \@ifxundefined [1]{%
 \@ifx{#1\undefined}
}%
\providecommand \@ifnum [1]{%
 \ifnum #1\expandafter \@firstoftwo
 \else \expandafter \@secondoftwo
 \fi
}%
\providecommand \@ifx [1]{%
 \ifx #1\expandafter \@firstoftwo
 \else \expandafter \@secondoftwo
 \fi
}%
\providecommand \natexlab [1]{#1}%
\providecommand \enquote  [1]{``#1''}%
\providecommand \bibnamefont  [1]{#1}%
\providecommand \bibfnamefont [1]{#1}%
\providecommand \citenamefont [1]{#1}%
\providecommand \href@noop [0]{\@secondoftwo}%
\providecommand \href [0]{\begingroup \@sanitize@url \@href}%
\providecommand \@href[1]{\@@startlink{#1}\@@href}%
\providecommand \@@href[1]{\endgroup#1\@@endlink}%
\providecommand \@sanitize@url [0]{\catcode `\\12\catcode `\$12\catcode
  `\&12\catcode `\#12\catcode `\^12\catcode `\_12\catcode `\%12\relax}%
\providecommand \@@startlink[1]{}%
\providecommand \@@endlink[0]{}%
\providecommand \url  [0]{\begingroup\@sanitize@url \@url }%
\providecommand \@url [1]{\endgroup\@href {#1}{\urlprefix }}%
\providecommand \urlprefix  [0]{URL }%
\providecommand \Eprint [0]{\href }%
\providecommand \doibase [0]{https://doi.org/}%
\providecommand \selectlanguage [0]{\@gobble}%
\providecommand \bibinfo  [0]{\@secondoftwo}%
\providecommand \bibfield  [0]{\@secondoftwo}%
\providecommand \translation [1]{[#1]}%
\providecommand \BibitemOpen [0]{}%
\providecommand \bibitemStop [0]{}%
\providecommand \bibitemNoStop [0]{.\EOS\space}%
\providecommand \EOS [0]{\spacefactor3000\relax}%
\providecommand \BibitemShut  [1]{\csname bibitem#1\endcsname}%
\let\auto@bib@innerbib\@empty
\bibitem [{\citenamefont {Vinjanampathy}\ and\ \citenamefont
  {Anders}(2016)}]{vinjanampathyQuantumThermodynamics2016}%
  \BibitemOpen
  \bibfield  {author} {\bibinfo {author} {\bibfnamefont {S.}~\bibnamefont
  {Vinjanampathy}}\ and\ \bibinfo {author} {\bibfnamefont {J.}~\bibnamefont
  {Anders}},\ }\bibfield  {title} {\bibinfo {title} {Quantum thermodynamics},\
  }\href {https://doi.org/10.1080/00107514.2016.1201896} {\bibfield  {journal}
  {\bibinfo  {journal} {Contemporary Physics}\ }\textbf {\bibinfo {volume}
  {57}},\ \bibinfo {pages} {545} (\bibinfo {year} {2016})}\BibitemShut
  {NoStop}%
\bibitem [{\citenamefont {Deffner}\ and\ \citenamefont
  {Campbell}(2019)}]{deffnerQuantumThermodynamicsIntroduction2019}%
  \BibitemOpen
  \bibfield  {author} {\bibinfo {author} {\bibfnamefont {S.}~\bibnamefont
  {Deffner}}\ and\ \bibinfo {author} {\bibfnamefont {S.}~\bibnamefont
  {Campbell}},\ }\bibfield  {title} {\bibinfo {title} {Quantum
  {{Thermodynamics}}: {{An}} introduction to the thermodynamics of quantum
  information},\ }\href@noop {} {\bibfield  {journal} {\bibinfo  {journal}
  {arXiv:1907.01596 [cond-mat, physics:quant-ph]}\ } (\bibinfo {year}
  {2019})},\ \Eprint {https://arxiv.org/abs/1907.01596} {arXiv:1907.01596
  [cond-mat, physics:quant-ph]} \BibitemShut {NoStop}%
\bibitem [{\citenamefont {Landi}\ and\ \citenamefont
  {Paternostro}(2021)}]{landiIrreversibleEntropyProduction2021}%
  \BibitemOpen
  \bibfield  {author} {\bibinfo {author} {\bibfnamefont {G.~T.}\ \bibnamefont
  {Landi}}\ and\ \bibinfo {author} {\bibfnamefont {M.}~\bibnamefont
  {Paternostro}},\ }\bibfield  {title} {\bibinfo {title} {Irreversible entropy
  production: {{From}} classical to quantum},\ }\href@noop {} {\bibfield
  {journal} {\bibinfo  {journal} {Reviews of Modern Physics}\ }\textbf
  {\bibinfo {volume} {93}},\ \bibinfo {pages} {035008} (\bibinfo {year}
  {2021})}\BibitemShut {NoStop}%
\bibitem [{\citenamefont {Esposito}\ \emph {et~al.}(2010)\citenamefont
  {Esposito}, \citenamefont {Lindenberg},\ and\ \citenamefont {den
  Broeck}}]{espositoEntropyProductionCorrelation2010}%
  \BibitemOpen
  \bibfield  {author} {\bibinfo {author} {\bibfnamefont {M.}~\bibnamefont
  {Esposito}}, \bibinfo {author} {\bibfnamefont {K.}~\bibnamefont
  {Lindenberg}},\ and\ \bibinfo {author} {\bibfnamefont {C.~V.}\ \bibnamefont
  {den Broeck}},\ }\bibfield  {title} {\bibinfo {title} {Entropy production as
  correlation between system and reservoir},\ }\href
  {https://doi.org/10.1088/1367-2630/12/1/013013} {\bibfield  {journal}
  {\bibinfo  {journal} {New Journal of Physics}\ }\textbf {\bibinfo {volume}
  {12}},\ \bibinfo {pages} {013013} (\bibinfo {year} {2010})}\BibitemShut
  {NoStop}%
\bibitem [{\citenamefont {Evans}\ and\ \citenamefont
  {Searles}(2002)}]{evansFluctuationTheorem2002}%
  \BibitemOpen
  \bibfield  {author} {\bibinfo {author} {\bibfnamefont {D.~J.}\ \bibnamefont
  {Evans}}\ and\ \bibinfo {author} {\bibfnamefont {D.~J.}\ \bibnamefont
  {Searles}},\ }\bibfield  {title} {\bibinfo {title} {The {{Fluctuation
  Theorem}}},\ }\href {https://doi.org/10.1080/00018730210155133} {\bibfield
  {journal} {\bibinfo  {journal} {Advances in Physics}\ }\textbf {\bibinfo
  {volume} {51}},\ \bibinfo {pages} {1529} (\bibinfo {year}
  {2002})}\BibitemShut {NoStop}%
\bibitem [{\citenamefont
  {Jarzynski}(2011)}]{jarzynskiEqualitiesInequalitiesIrreversibility2011}%
  \BibitemOpen
  \bibfield  {author} {\bibinfo {author} {\bibfnamefont {C.}~\bibnamefont
  {Jarzynski}},\ }\bibfield  {title} {\bibinfo {title} {Equalities and
  {{Inequalities}}: {{Irreversibility}} and the {{Second Law}} of
  {{Thermodynamics}} at the {{Nanoscale}}},\ }\href
  {https://doi.org/10.1146/annurev-conmatphys-062910-140506} {\bibfield
  {journal} {\bibinfo  {journal} {Annual Review of Condensed Matter Physics}\
  }\textbf {\bibinfo {volume} {2}},\ \bibinfo {pages} {329} (\bibinfo {year}
  {2011})}\BibitemShut {NoStop}%
\bibitem [{\citenamefont
  {Seifert}(2012)}]{seifertStochasticThermodynamicsFluctuation2012}%
  \BibitemOpen
  \bibfield  {author} {\bibinfo {author} {\bibfnamefont {U.}~\bibnamefont
  {Seifert}},\ }\bibfield  {title} {\bibinfo {title} {Stochastic
  thermodynamics, fluctuation theorems and molecular machines},\ }\href
  {https://doi.org/10.1088/0034-4885/75/12/126001} {\bibfield  {journal}
  {\bibinfo  {journal} {Reports on Progress in Physics}\ }\textbf {\bibinfo
  {volume} {75}},\ \bibinfo {pages} {126001} (\bibinfo {year}
  {2012})}\BibitemShut {NoStop}%
\bibitem [{\citenamefont {Tasaki}(2000)}]{tasakiJarzynskiRelationsQuantum2000}%
  \BibitemOpen
  \bibfield  {author} {\bibinfo {author} {\bibfnamefont {H.}~\bibnamefont
  {Tasaki}},\ }\bibfield  {title} {\bibinfo {title} {Jarzynski relations for
  quantum systems and some applications},\ }\href@noop {} {\bibfield  {journal}
  {\bibinfo  {journal} {arXiv preprint cond-mat/0009244}\ } (\bibinfo {year}
  {2000})},\ \Eprint {https://arxiv.org/abs/cond-mat/0009244}
  {arXiv:cond-mat/0009244} \BibitemShut {NoStop}%
\bibitem [{\citenamefont {Esposito}\ \emph {et~al.}(2009)\citenamefont
  {Esposito}, \citenamefont {Harbola},\ and\ \citenamefont
  {Mukamel}}]{espositoNonequilibriumFluctuationsFluctuation2009}%
  \BibitemOpen
  \bibfield  {author} {\bibinfo {author} {\bibfnamefont {M.}~\bibnamefont
  {Esposito}}, \bibinfo {author} {\bibfnamefont {U.}~\bibnamefont {Harbola}},\
  and\ \bibinfo {author} {\bibfnamefont {S.}~\bibnamefont {Mukamel}},\
  }\bibfield  {title} {\bibinfo {title} {Nonequilibrium fluctuations,
  fluctuation theorems, and counting statistics in quantum systems},\ }\href
  {https://doi.org/10.1103/RevModPhys.81.1665} {\bibfield  {journal} {\bibinfo
  {journal} {Reviews of Modern Physics}\ }\textbf {\bibinfo {volume} {81}},\
  \bibinfo {pages} {1665} (\bibinfo {year} {2009})}\BibitemShut {NoStop}%
\bibitem [{\citenamefont {Campisi}\ \emph {et~al.}(2011)\citenamefont
  {Campisi}, \citenamefont {H{\"a}nggi},\ and\ \citenamefont
  {Talkner}}]{campisiColloquiumQuantumFluctuation2011}%
  \BibitemOpen
  \bibfield  {author} {\bibinfo {author} {\bibfnamefont {M.}~\bibnamefont
  {Campisi}}, \bibinfo {author} {\bibfnamefont {P.}~\bibnamefont
  {H{\"a}nggi}},\ and\ \bibinfo {author} {\bibfnamefont {P.}~\bibnamefont
  {Talkner}},\ }\bibfield  {title} {\bibinfo {title} {Colloquium: {{Quantum}}
  fluctuation relations: {{Foundations}} and applications},\ }\href
  {https://doi.org/10.1103/RevModPhys.83.771} {\bibfield  {journal} {\bibinfo
  {journal} {Reviews of Modern Physics}\ }\textbf {\bibinfo {volume} {83}},\
  \bibinfo {pages} {771} (\bibinfo {year} {2011})}\BibitemShut {NoStop}%
\bibitem [{\citenamefont {Funo}\ \emph {et~al.}(2018)\citenamefont {Funo},
  \citenamefont {Ueda},\ and\ \citenamefont
  {Sagawa}}]{funoQuantumFluctuationTheorems2018}%
  \BibitemOpen
  \bibfield  {author} {\bibinfo {author} {\bibfnamefont {K.}~\bibnamefont
  {Funo}}, \bibinfo {author} {\bibfnamefont {M.}~\bibnamefont {Ueda}},\ and\
  \bibinfo {author} {\bibfnamefont {T.}~\bibnamefont {Sagawa}},\ }\bibfield
  {title} {\bibinfo {title} {Quantum {{Fluctuation Theorems}}},\ }in\ \href
  {https://doi.org/10.1007/978-3-319-99046-0_10} {\emph {\bibinfo {booktitle}
  {Thermodynamics in the {{Quantum Regime}}: {{Fundamental Aspects}} and {{New
  Directions}}}}},\ \bibinfo {series and number} {Fundamental {{Theories}} of
  {{Physics}}},\ \bibinfo {editor} {edited by\ \bibinfo {editor} {\bibfnamefont
  {F.}~\bibnamefont {Binder}}, \bibinfo {editor} {\bibfnamefont {L.~A.}\
  \bibnamefont {Correa}}, \bibinfo {editor} {\bibfnamefont {C.}~\bibnamefont
  {Gogolin}}, \bibinfo {editor} {\bibfnamefont {J.}~\bibnamefont {Anders}},\
  and\ \bibinfo {editor} {\bibfnamefont {G.}~\bibnamefont {Adesso}}}\ (\bibinfo
   {publisher} {{Springer International Publishing}},\ \bibinfo {address}
  {{Cham}},\ \bibinfo {year} {2018})\ pp.\ \bibinfo {pages}
  {249--273}\BibitemShut {NoStop}%
\bibitem [{\citenamefont {Maruyama}\ \emph {et~al.}(2009)\citenamefont
  {Maruyama}, \citenamefont {Nori},\ and\ \citenamefont
  {Vedral}}]{maruyamaColloquiumPhysicsMaxwell2009}%
  \BibitemOpen
  \bibfield  {author} {\bibinfo {author} {\bibfnamefont {K.}~\bibnamefont
  {Maruyama}}, \bibinfo {author} {\bibfnamefont {F.}~\bibnamefont {Nori}},\
  and\ \bibinfo {author} {\bibfnamefont {V.}~\bibnamefont {Vedral}},\
  }\bibfield  {title} {\bibinfo {title} {Colloquium: {{The}} physics of
  {{Maxwell}}'s demon and information},\ }\href
  {https://doi.org/10.1103/RevModPhys.81.1} {\bibfield  {journal} {\bibinfo
  {journal} {Reviews of Modern Physics}\ }\textbf {\bibinfo {volume} {81}},\
  \bibinfo {pages} {1} (\bibinfo {year} {2009})}\BibitemShut {NoStop}%
\bibitem [{\citenamefont {Sagawa}\ and\ \citenamefont
  {Ueda}(2008)}]{sagawaSecondLawThermodynamics2008}%
  \BibitemOpen
  \bibfield  {author} {\bibinfo {author} {\bibfnamefont {T.}~\bibnamefont
  {Sagawa}}\ and\ \bibinfo {author} {\bibfnamefont {M.}~\bibnamefont {Ueda}},\
  }\bibfield  {title} {\bibinfo {title} {Second {{Law}} of {{Thermodynamics}}
  with {{Discrete Quantum Feedback Control}}},\ }\href
  {https://doi.org/10.1103/PhysRevLett.100.080403} {\bibfield  {journal}
  {\bibinfo  {journal} {Physical Review Letters}\ }\textbf {\bibinfo {volume}
  {100}},\ \bibinfo {pages} {080403} (\bibinfo {year} {2008})}\BibitemShut
  {NoStop}%
\bibitem [{\citenamefont
  {Sagawa}(2012)}]{sagawaThermodynamicsInformationProcessing2012}%
  \BibitemOpen
  \bibfield  {author} {\bibinfo {author} {\bibfnamefont {T.}~\bibnamefont
  {Sagawa}},\ }\bibfield  {title} {\bibinfo {title} {Thermodynamics of
  {{Information Processing}} in {{Small Systems}}*)},\ }\href
  {https://doi.org/10.1143/PTP.127.1} {\bibfield  {journal} {\bibinfo
  {journal} {Progress of Theoretical Physics}\ }\textbf {\bibinfo {volume}
  {127}},\ \bibinfo {pages} {1} (\bibinfo {year} {2012})}\BibitemShut {NoStop}%
\bibitem [{\citenamefont {Funo}\ \emph {et~al.}(2015)\citenamefont {Funo},
  \citenamefont {Murashita},\ and\ \citenamefont
  {Ueda}}]{funoQuantumNonequilibriumEqualities2015}%
  \BibitemOpen
  \bibfield  {author} {\bibinfo {author} {\bibfnamefont {K.}~\bibnamefont
  {Funo}}, \bibinfo {author} {\bibfnamefont {Y.}~\bibnamefont {Murashita}},\
  and\ \bibinfo {author} {\bibfnamefont {M.}~\bibnamefont {Ueda}},\ }\bibfield
  {title} {\bibinfo {title} {Quantum nonequilibrium equalities with absolute
  irreversibility},\ }\href {https://doi.org/10.1088/1367-2630/17/7/075005}
  {\bibfield  {journal} {\bibinfo  {journal} {New Journal of Physics}\ }\textbf
  {\bibinfo {volume} {17}},\ \bibinfo {pages} {075005} (\bibinfo {year}
  {2015})}\BibitemShut {NoStop}%
\bibitem [{\citenamefont {Camati}\ \emph {et~al.}(2016)\citenamefont {Camati},
  \citenamefont {Peterson}, \citenamefont {Batalh{\~a}o}, \citenamefont
  {Micadei}, \citenamefont {Souza}, \citenamefont {Sarthour}, \citenamefont
  {Oliveira},\ and\ \citenamefont
  {Serra}}]{camatiExperimentalRectificationEntropy2016}%
  \BibitemOpen
  \bibfield  {author} {\bibinfo {author} {\bibfnamefont {P.~A.}\ \bibnamefont
  {Camati}}, \bibinfo {author} {\bibfnamefont {J.~P.~S.}\ \bibnamefont
  {Peterson}}, \bibinfo {author} {\bibfnamefont {T.~B.}\ \bibnamefont
  {Batalh{\~a}o}}, \bibinfo {author} {\bibfnamefont {K.}~\bibnamefont
  {Micadei}}, \bibinfo {author} {\bibfnamefont {A.~M.}\ \bibnamefont {Souza}},
  \bibinfo {author} {\bibfnamefont {R.~S.}\ \bibnamefont {Sarthour}}, \bibinfo
  {author} {\bibfnamefont {I.~S.}\ \bibnamefont {Oliveira}},\ and\ \bibinfo
  {author} {\bibfnamefont {R.~M.}\ \bibnamefont {Serra}},\ }\bibfield  {title}
  {\bibinfo {title} {Experimental {{Rectification}} of {{Entropy Production}}
  by {{Maxwell}}'s {{Demon}} in a {{Quantum System}}},\ }\href
  {https://doi.org/10.1103/PhysRevLett.117.240502} {\bibfield  {journal}
  {\bibinfo  {journal} {Physical Review Letters}\ }\textbf {\bibinfo {volume}
  {117}},\ \bibinfo {pages} {240502} (\bibinfo {year} {2016})}\BibitemShut
  {NoStop}%
\bibitem [{\citenamefont {Naghiloo}\ \emph {et~al.}(2018)\citenamefont
  {Naghiloo}, \citenamefont {Alonso}, \citenamefont {Romito}, \citenamefont
  {Lutz},\ and\ \citenamefont {Murch}}]{naghilooInformationGainLoss2018}%
  \BibitemOpen
  \bibfield  {author} {\bibinfo {author} {\bibfnamefont {M.}~\bibnamefont
  {Naghiloo}}, \bibinfo {author} {\bibfnamefont {J.~J.}\ \bibnamefont
  {Alonso}}, \bibinfo {author} {\bibfnamefont {A.}~\bibnamefont {Romito}},
  \bibinfo {author} {\bibfnamefont {E.}~\bibnamefont {Lutz}},\ and\ \bibinfo
  {author} {\bibfnamefont {K.~W.}\ \bibnamefont {Murch}},\ }\bibfield  {title}
  {\bibinfo {title} {Information {{Gain}} and {{Loss}} for a {{Quantum
  Maxwell}}'s {{Demon}}},\ }\href
  {https://doi.org/10.1103/PhysRevLett.121.030604} {\bibfield  {journal}
  {\bibinfo  {journal} {Physical Review Letters}\ }\textbf {\bibinfo {volume}
  {121}},\ \bibinfo {pages} {030604} (\bibinfo {year} {2018})}\BibitemShut
  {NoStop}%
\bibitem [{\citenamefont {Tan}\ \emph {et~al.}(2021)\citenamefont {Tan},
  \citenamefont {Camati}, \citenamefont {Cauquil}, \citenamefont
  {Auff{\`e}ves},\ and\ \citenamefont
  {Dotsenko}}]{tanAlternativeExperimentalWays2021}%
  \BibitemOpen
  \bibfield  {author} {\bibinfo {author} {\bibfnamefont {Z.}~\bibnamefont
  {Tan}}, \bibinfo {author} {\bibfnamefont {P.~A.}\ \bibnamefont {Camati}},
  \bibinfo {author} {\bibfnamefont {G.~C.}\ \bibnamefont {Cauquil}}, \bibinfo
  {author} {\bibfnamefont {A.}~\bibnamefont {Auff{\`e}ves}},\ and\ \bibinfo
  {author} {\bibfnamefont {I.}~\bibnamefont {Dotsenko}},\ }\bibfield  {title}
  {\bibinfo {title} {Alternative experimental ways to access entropy
  production},\ }\href {https://doi.org/10.1103/PhysRevResearch.3.043076}
  {\bibfield  {journal} {\bibinfo  {journal} {Physical Review Research}\
  }\textbf {\bibinfo {volume} {3}},\ \bibinfo {pages} {043076} (\bibinfo {year}
  {2021})}\BibitemShut {NoStop}%
\bibitem [{\citenamefont {Lloyd}(1989)}]{lloydUseMutualInformation1989}%
  \BibitemOpen
  \bibfield  {author} {\bibinfo {author} {\bibfnamefont {S.}~\bibnamefont
  {Lloyd}},\ }\bibfield  {title} {\bibinfo {title} {Use of mutual information
  to decrease entropy: {{Implications}} for the second law of thermodynamics},\
  }\href {https://doi.org/10.1103/PhysRevA.39.5378} {\bibfield  {journal}
  {\bibinfo  {journal} {Physical Review A}\ }\textbf {\bibinfo {volume} {39}},\
  \bibinfo {pages} {5378} (\bibinfo {year} {1989})}\BibitemShut {NoStop}%
\bibitem [{\citenamefont {Cerf}\ and\ \citenamefont
  {Adami}(1997)}]{cerfNegativeEntropyInformation1997}%
  \BibitemOpen
  \bibfield  {author} {\bibinfo {author} {\bibfnamefont {N.~J.}\ \bibnamefont
  {Cerf}}\ and\ \bibinfo {author} {\bibfnamefont {C.}~\bibnamefont {Adami}},\
  }\bibfield  {title} {\bibinfo {title} {Negative {{Entropy}} and
  {{Information}} in {{Quantum Mechanics}}},\ }\href
  {https://doi.org/10.1103/PhysRevLett.79.5194} {\bibfield  {journal} {\bibinfo
   {journal} {Physical Review Letters}\ }\textbf {\bibinfo {volume} {79}},\
  \bibinfo {pages} {5194} (\bibinfo {year} {1997})}\BibitemShut {NoStop}%
\bibitem [{\citenamefont {Horodecki}\ \emph {et~al.}(2005)\citenamefont
  {Horodecki}, \citenamefont {Oppenheim},\ and\ \citenamefont
  {Winter}}]{horodeckiPartialQuantumInformation2005}%
  \BibitemOpen
  \bibfield  {author} {\bibinfo {author} {\bibfnamefont {M.}~\bibnamefont
  {Horodecki}}, \bibinfo {author} {\bibfnamefont {J.}~\bibnamefont
  {Oppenheim}},\ and\ \bibinfo {author} {\bibfnamefont {A.}~\bibnamefont
  {Winter}},\ }\bibfield  {title} {\bibinfo {title} {Partial quantum
  information},\ }\href {https://doi.org/10.1038/nature03909} {\bibfield
  {journal} {\bibinfo  {journal} {Nature}\ }\textbf {\bibinfo {volume} {436}},\
  \bibinfo {pages} {673} (\bibinfo {year} {2005})}\BibitemShut {NoStop}%
\bibitem [{\citenamefont {del Rio}\ \emph {et~al.}(2011)\citenamefont {del
  Rio}, \citenamefont {{\AA}berg}, \citenamefont {Renner}, \citenamefont
  {Dahlsten},\ and\ \citenamefont
  {Vedral}}]{rioThermodynamicMeaningNegative2011}%
  \BibitemOpen
  \bibfield  {author} {\bibinfo {author} {\bibfnamefont {L.}~\bibnamefont {del
  Rio}}, \bibinfo {author} {\bibfnamefont {J.}~\bibnamefont {{\AA}berg}},
  \bibinfo {author} {\bibfnamefont {R.}~\bibnamefont {Renner}}, \bibinfo
  {author} {\bibfnamefont {O.}~\bibnamefont {Dahlsten}},\ and\ \bibinfo
  {author} {\bibfnamefont {V.}~\bibnamefont {Vedral}},\ }\bibfield  {title}
  {\bibinfo {title} {The thermodynamic meaning of negative entropy},\ }\href
  {https://doi.org/10.1038/nature10123} {\bibfield  {journal} {\bibinfo
  {journal} {Nature}\ }\textbf {\bibinfo {volume} {474}},\ \bibinfo {pages}
  {61} (\bibinfo {year} {2011})}\BibitemShut {NoStop}%
\bibitem [{\citenamefont {Belenchia}\ \emph {et~al.}(2020)\citenamefont
  {Belenchia}, \citenamefont {Mancino}, \citenamefont {Landi},\ and\
  \citenamefont {Paternostro}}]{belenchiaEntropyProductionContinuously2020}%
  \BibitemOpen
  \bibfield  {author} {\bibinfo {author} {\bibfnamefont {A.}~\bibnamefont
  {Belenchia}}, \bibinfo {author} {\bibfnamefont {L.}~\bibnamefont {Mancino}},
  \bibinfo {author} {\bibfnamefont {G.~T.}\ \bibnamefont {Landi}},\ and\
  \bibinfo {author} {\bibfnamefont {M.}~\bibnamefont {Paternostro}},\
  }\bibfield  {title} {\bibinfo {title} {Entropy production in continuously
  measured {{Gaussian}} quantum systems},\ }\href
  {https://doi.org/10.1038/s41534-020-00334-6} {\bibfield  {journal} {\bibinfo
  {journal} {npj Quantum Information}\ }\textbf {\bibinfo {volume} {6}},\
  \bibinfo {pages} {1} (\bibinfo {year} {2020})}\BibitemShut {NoStop}%
\bibitem [{\citenamefont {Landi}\ \emph {et~al.}(2022)\citenamefont {Landi},
  \citenamefont {Paternostro},\ and\ \citenamefont
  {Belenchia}}]{landiInformationalSteadyStates2022}%
  \BibitemOpen
  \bibfield  {author} {\bibinfo {author} {\bibfnamefont {G.~T.}\ \bibnamefont
  {Landi}}, \bibinfo {author} {\bibfnamefont {M.}~\bibnamefont {Paternostro}},\
  and\ \bibinfo {author} {\bibfnamefont {A.}~\bibnamefont {Belenchia}},\
  }\bibfield  {title} {\bibinfo {title} {Informational {{Steady States}} and
  {{Conditional Entropy Production}} in {{Continuously Monitored Systems}}},\
  }\href {https://doi.org/10.1103/PRXQuantum.3.010303} {\bibfield  {journal}
  {\bibinfo  {journal} {PRX Quantum}\ }\textbf {\bibinfo {volume} {3}},\
  \bibinfo {pages} {010303} (\bibinfo {year} {2022})}\BibitemShut {NoStop}%
\bibitem [{\citenamefont {Funo}\ \emph {et~al.}(2013)\citenamefont {Funo},
  \citenamefont {Watanabe},\ and\ \citenamefont
  {Ueda}}]{funoThermodynamicWorkGain2013}%
  \BibitemOpen
  \bibfield  {author} {\bibinfo {author} {\bibfnamefont {K.}~\bibnamefont
  {Funo}}, \bibinfo {author} {\bibfnamefont {Y.}~\bibnamefont {Watanabe}},\
  and\ \bibinfo {author} {\bibfnamefont {M.}~\bibnamefont {Ueda}},\ }\bibfield
  {title} {\bibinfo {title} {Thermodynamic work gain from entanglement},\
  }\href {https://doi.org/10.1103/PhysRevA.88.052319} {\bibfield  {journal}
  {\bibinfo  {journal} {Physical Review A}\ }\textbf {\bibinfo {volume} {88}},\
  \bibinfo {pages} {052319} (\bibinfo {year} {2013})}\BibitemShut {NoStop}%
\bibitem [{\citenamefont {Bera}\ \emph {et~al.}(2017)\citenamefont {Bera},
  \citenamefont {Riera}, \citenamefont {Lewenstein},\ and\ \citenamefont
  {Winter}}]{beraGeneralizedLawsThermodynamics2017}%
  \BibitemOpen
  \bibfield  {author} {\bibinfo {author} {\bibfnamefont {M.~N.}\ \bibnamefont
  {Bera}}, \bibinfo {author} {\bibfnamefont {A.}~\bibnamefont {Riera}},
  \bibinfo {author} {\bibfnamefont {M.}~\bibnamefont {Lewenstein}},\ and\
  \bibinfo {author} {\bibfnamefont {A.}~\bibnamefont {Winter}},\ }\bibfield
  {title} {\bibinfo {title} {Generalized laws of thermodynamics in the presence
  of correlations},\ }\href {https://doi.org/10.1038/s41467-017-02370-x}
  {\bibfield  {journal} {\bibinfo  {journal} {Nature Communications}\ }\textbf
  {\bibinfo {volume} {8}},\ \bibinfo {pages} {2180} (\bibinfo {year}
  {2017})}\BibitemShut {NoStop}%
\bibitem [{\citenamefont {Manzano}\ \emph
  {et~al.}(2018{\natexlab{a}})\citenamefont {Manzano}, \citenamefont
  {Plastina},\ and\ \citenamefont
  {Zambrini}}]{manzanoOptimalWorkExtraction2018}%
  \BibitemOpen
  \bibfield  {author} {\bibinfo {author} {\bibfnamefont {G.}~\bibnamefont
  {Manzano}}, \bibinfo {author} {\bibfnamefont {F.}~\bibnamefont {Plastina}},\
  and\ \bibinfo {author} {\bibfnamefont {R.}~\bibnamefont {Zambrini}},\
  }\bibfield  {title} {\bibinfo {title} {Optimal {{Work Extraction}} and
  {{Thermodynamics}} of {{Quantum Measurements}} and {{Correlations}}},\ }\href
  {https://doi.org/10.1103/PhysRevLett.121.120602} {\bibfield  {journal}
  {\bibinfo  {journal} {Physical Review Letters}\ }\textbf {\bibinfo {volume}
  {121}},\ \bibinfo {pages} {120602} (\bibinfo {year}
  {2018}{\natexlab{a}})}\BibitemShut {NoStop}%
\bibitem [{\citenamefont {Jarzynski}\ and\ \citenamefont
  {W{\'o}jcik}(2004)}]{jarzynskiClassicalQuantumFluctuation2004}%
  \BibitemOpen
  \bibfield  {author} {\bibinfo {author} {\bibfnamefont {C.}~\bibnamefont
  {Jarzynski}}\ and\ \bibinfo {author} {\bibfnamefont {D.~K.}\ \bibnamefont
  {W{\'o}jcik}},\ }\bibfield  {title} {\bibinfo {title} {Classical and
  {{Quantum Fluctuation Theorems}} for {{Heat Exchange}}},\ }\href
  {https://doi.org/10.1103/PhysRevLett.92.230602} {\bibfield  {journal}
  {\bibinfo  {journal} {Physical Review Letters}\ }\textbf {\bibinfo {volume}
  {92}},\ \bibinfo {pages} {230602} (\bibinfo {year} {2004})}\BibitemShut
  {NoStop}%
\bibitem [{\citenamefont {Manzano}\ \emph
  {et~al.}(2018{\natexlab{b}})\citenamefont {Manzano}, \citenamefont
  {Horowitz},\ and\ \citenamefont
  {Parrondo}}]{manzanoQuantumFluctuationTheorems2018}%
  \BibitemOpen
  \bibfield  {author} {\bibinfo {author} {\bibfnamefont {G.}~\bibnamefont
  {Manzano}}, \bibinfo {author} {\bibfnamefont {J.~M.}\ \bibnamefont
  {Horowitz}},\ and\ \bibinfo {author} {\bibfnamefont {J.~M.~R.}\ \bibnamefont
  {Parrondo}},\ }\bibfield  {title} {\bibinfo {title} {Quantum {{Fluctuation
  Theorems}} for {{Arbitrary Environments}}: {{Adiabatic}} and {{Nonadiabatic
  Entropy Production}}},\ }\href {https://doi.org/10.1103/PhysRevX.8.031037}
  {\bibfield  {journal} {\bibinfo  {journal} {Physical Review X}\ }\textbf
  {\bibinfo {volume} {8}},\ \bibinfo {pages} {031037} (\bibinfo {year}
  {2018}{\natexlab{b}})}\BibitemShut {NoStop}%
\bibitem [{\citenamefont {Nielsen}\ and\ \citenamefont
  {Chuang}(2010)}]{nielsenQuantumComputationQuantum2010}%
  \BibitemOpen
  \bibfield  {author} {\bibinfo {author} {\bibfnamefont {M.~A.}\ \bibnamefont
  {Nielsen}}\ and\ \bibinfo {author} {\bibfnamefont {I.~L.}\ \bibnamefont
  {Chuang}},\ }\href {https://doi.org/10.1017/CBO9780511976667} {\bibinfo
  {title} {Quantum {{Computation}} and {{Quantum Information}}: 10th
  {{Anniversary Edition}}}} (\bibinfo {year} {2010})\BibitemShut {NoStop}%
\bibitem [{\citenamefont {Linden}\ \emph {et~al.}(2009)\citenamefont {Linden},
  \citenamefont {Popescu}, \citenamefont {Short},\ and\ \citenamefont
  {Winter}}]{lindenQuantumMechanicalEvolution2009}%
  \BibitemOpen
  \bibfield  {author} {\bibinfo {author} {\bibfnamefont {N.}~\bibnamefont
  {Linden}}, \bibinfo {author} {\bibfnamefont {S.}~\bibnamefont {Popescu}},
  \bibinfo {author} {\bibfnamefont {A.~J.}\ \bibnamefont {Short}},\ and\
  \bibinfo {author} {\bibfnamefont {A.}~\bibnamefont {Winter}},\ }\bibfield
  {title} {\bibinfo {title} {Quantum mechanical evolution towards thermal
  equilibrium},\ }\href {https://doi.org/10.1103/PhysRevE.79.061103} {\bibfield
   {journal} {\bibinfo  {journal} {Physical Review E}\ }\textbf {\bibinfo
  {volume} {79}},\ \bibinfo {pages} {061103} (\bibinfo {year}
  {2009})}\BibitemShut {NoStop}%
\bibitem [{\citenamefont {Lloyd}(2013)}]{lloydPureStateQuantum2013}%
  \BibitemOpen
  \bibfield  {author} {\bibinfo {author} {\bibfnamefont {S.}~\bibnamefont
  {Lloyd}},\ }\bibfield  {title} {\bibinfo {title} {Pure state quantum
  statistical mechanics and black holes},\ }\href@noop {} {\bibfield  {journal}
  {\bibinfo  {journal} {arXiv:1307.0378 [quant-ph]}\ } (\bibinfo {year}
  {2013})},\ \Eprint {https://arxiv.org/abs/1307.0378} {arXiv:1307.0378
  [quant-ph]} \BibitemShut {NoStop}%
\bibitem [{\citenamefont {Wilde}(2017)}]{wildeClassicalQuantumShannon2017}%
  \BibitemOpen
  \bibfield  {author} {\bibinfo {author} {\bibfnamefont {M.~M.}\ \bibnamefont
  {Wilde}},\ }\bibfield  {title} {\bibinfo {title} {From {{Classical}} to
  {{Quantum Shannon Theory}}},\ }\bibfield  {journal} {\bibinfo  {journal}
  {arXiv:1106.1445 [quant-ph]}\ }\href
  {https://doi.org/10.1017/9781316809976.001} {10.1017/9781316809976.001}
  (\bibinfo {year} {2017}),\ \Eprint {https://arxiv.org/abs/1106.1445}
  {arXiv:1106.1445 [quant-ph]} \BibitemShut {NoStop}%
\bibitem [{\citenamefont {Schumacher}\ and\ \citenamefont
  {Nielsen}(1996)}]{schumacherQuantumDataProcessing1996}%
  \BibitemOpen
  \bibfield  {author} {\bibinfo {author} {\bibfnamefont {B.}~\bibnamefont
  {Schumacher}}\ and\ \bibinfo {author} {\bibfnamefont {M.~A.}\ \bibnamefont
  {Nielsen}},\ }\bibfield  {title} {\bibinfo {title} {Quantum data processing
  and error correction},\ }\href {https://doi.org/10.1103/PhysRevA.54.2629}
  {\bibfield  {journal} {\bibinfo  {journal} {Physical Review A}\ }\textbf
  {\bibinfo {volume} {54}},\ \bibinfo {pages} {2629} (\bibinfo {year}
  {1996})}\BibitemShut {NoStop}%
\bibitem [{\citenamefont {Lieb}\ and\ \citenamefont
  {Ruskai}(1973)}]{liebProofStrongSubadditivity1973}%
  \BibitemOpen
  \bibfield  {author} {\bibinfo {author} {\bibfnamefont {E.~H.}\ \bibnamefont
  {Lieb}}\ and\ \bibinfo {author} {\bibfnamefont {M.~B.}\ \bibnamefont
  {Ruskai}},\ }\bibfield  {title} {\bibinfo {title} {Proof of the strong
  subadditivity of quantum-mechanical entropy},\ }\href
  {https://doi.org/10.1063/1.1666274} {\bibfield  {journal} {\bibinfo
  {journal} {Journal of Mathematical Physics}\ }\textbf {\bibinfo {volume}
  {14}},\ \bibinfo {pages} {1938} (\bibinfo {year} {1973})}\BibitemShut
  {NoStop}%
\bibitem [{\citenamefont {Zeng}\ and\ \citenamefont
  {Wang}(2021)}]{zengNewFluctuationTheorems}%
  \BibitemOpen
  \bibfield  {author} {\bibinfo {author} {\bibfnamefont {Q.}~\bibnamefont
  {Zeng}}\ and\ \bibinfo {author} {\bibfnamefont {J.}~\bibnamefont {Wang}},\
  }\bibfield  {title} {\bibinfo {title} {New fluctuation theorems on
  {{Maxwell}}'s demon},\ }\href {https://doi.org/10.1126/sciadv.abf1807}
  {\bibfield  {journal} {\bibinfo  {journal} {Science Advances}\ }\textbf
  {\bibinfo {volume} {7}},\ \bibinfo {pages} {eabf1807} (\bibinfo {year}
  {2021})}\BibitemShut {NoStop}%
\bibitem [{\citenamefont {Hayden}\ \emph {et~al.}(2004)\citenamefont {Hayden},
  \citenamefont {Jozsa}, \citenamefont {Petz},\ and\ \citenamefont
  {Winter}}]{haydenStructureStatesWhich2004}%
  \BibitemOpen
  \bibfield  {author} {\bibinfo {author} {\bibfnamefont {P.}~\bibnamefont
  {Hayden}}, \bibinfo {author} {\bibfnamefont {R.}~\bibnamefont {Jozsa}},
  \bibinfo {author} {\bibfnamefont {D.}~\bibnamefont {Petz}},\ and\ \bibinfo
  {author} {\bibfnamefont {A.}~\bibnamefont {Winter}},\ }\bibfield  {title}
  {\bibinfo {title} {Structure of {{States Which Satisfy Strong Subadditivity}}
  of {{Quantum Entropy}} with {{Equality}}},\ }\href
  {https://doi.org/10.1007/s00220-004-1049-z} {\bibfield  {journal} {\bibinfo
  {journal} {Communications in Mathematical Physics}\ }\textbf {\bibinfo
  {volume} {246}},\ \bibinfo {pages} {359} (\bibinfo {year}
  {2004})}\BibitemShut {NoStop}%
\bibitem [{\citenamefont {Cubitt}\ \emph {et~al.}(2003)\citenamefont {Cubitt},
  \citenamefont {Verstraete}, \citenamefont {D{\"u}r},\ and\ \citenamefont
  {Cirac}}]{cubittSeparableStatesCan2003}%
  \BibitemOpen
  \bibfield  {author} {\bibinfo {author} {\bibfnamefont {T.~S.}\ \bibnamefont
  {Cubitt}}, \bibinfo {author} {\bibfnamefont {F.}~\bibnamefont {Verstraete}},
  \bibinfo {author} {\bibfnamefont {W.}~\bibnamefont {D{\"u}r}},\ and\ \bibinfo
  {author} {\bibfnamefont {J.~I.}\ \bibnamefont {Cirac}},\ }\bibfield  {title}
  {\bibinfo {title} {Separable {{States Can Be Used To Distribute
  Entanglement}}},\ }\href {https://doi.org/10.1103/PhysRevLett.91.037902}
  {\bibfield  {journal} {\bibinfo  {journal} {Physical Review Letters}\
  }\textbf {\bibinfo {volume} {91}},\ \bibinfo {pages} {037902} (\bibinfo
  {year} {2003})}\BibitemShut {NoStop}%
\bibitem [{\citenamefont {Chuan}\ \emph {et~al.}(2012)\citenamefont {Chuan},
  \citenamefont {Maillard}, \citenamefont {Modi}, \citenamefont {Paterek},
  \citenamefont {Paternostro},\ and\ \citenamefont
  {Piani}}]{chuanQuantumDiscordBounds2012}%
  \BibitemOpen
  \bibfield  {author} {\bibinfo {author} {\bibfnamefont {T.~K.}\ \bibnamefont
  {Chuan}}, \bibinfo {author} {\bibfnamefont {J.}~\bibnamefont {Maillard}},
  \bibinfo {author} {\bibfnamefont {K.}~\bibnamefont {Modi}}, \bibinfo {author}
  {\bibfnamefont {T.}~\bibnamefont {Paterek}}, \bibinfo {author} {\bibfnamefont
  {M.}~\bibnamefont {Paternostro}},\ and\ \bibinfo {author} {\bibfnamefont
  {M.}~\bibnamefont {Piani}},\ }\bibfield  {title} {\bibinfo {title} {Quantum
  {{Discord Bounds}} the {{Amount}} of {{Distributed Entanglement}}},\ }\href
  {https://doi.org/10.1103/PhysRevLett.109.070501} {\bibfield  {journal}
  {\bibinfo  {journal} {Physical Review Letters}\ }\textbf {\bibinfo {volume}
  {109}},\ \bibinfo {pages} {070501} (\bibinfo {year} {2012})}\BibitemShut
  {NoStop}%
\bibitem [{\citenamefont {Fawzi}\ and\ \citenamefont
  {Renner}(2015)}]{fawziQuantumConditionalMutual2015}%
  \BibitemOpen
  \bibfield  {author} {\bibinfo {author} {\bibfnamefont {O.}~\bibnamefont
  {Fawzi}}\ and\ \bibinfo {author} {\bibfnamefont {R.}~\bibnamefont {Renner}},\
  }\bibfield  {title} {\bibinfo {title} {Quantum {{Conditional Mutual
  Information}} and {{Approximate Markov Chains}}},\ }\href
  {https://doi.org/10.1007/s00220-015-2466-x} {\bibfield  {journal} {\bibinfo
  {journal} {Communications in Mathematical Physics}\ }\textbf {\bibinfo
  {volume} {340}},\ \bibinfo {pages} {575} (\bibinfo {year}
  {2015})}\BibitemShut {NoStop}%
\bibitem [{\citenamefont {Brand{\~a}o}\ \emph {et~al.}(2015)\citenamefont
  {Brand{\~a}o}, \citenamefont {Harrow}, \citenamefont {Oppenheim},\ and\
  \citenamefont {Strelchuk}}]{brandaoQuantumConditionalMutual2015}%
  \BibitemOpen
  \bibfield  {author} {\bibinfo {author} {\bibfnamefont {F.~G. S.~L.}\
  \bibnamefont {Brand{\~a}o}}, \bibinfo {author} {\bibfnamefont {A.~W.}\
  \bibnamefont {Harrow}}, \bibinfo {author} {\bibfnamefont {J.}~\bibnamefont
  {Oppenheim}},\ and\ \bibinfo {author} {\bibfnamefont {S.}~\bibnamefont
  {Strelchuk}},\ }\bibfield  {title} {\bibinfo {title} {Quantum {{Conditional
  Mutual Information}}, {{Reconstructed States}}, and {{State
  Redistribution}}},\ }\href {https://doi.org/10.1103/PhysRevLett.115.050501}
  {\bibfield  {journal} {\bibinfo  {journal} {Physical Review Letters}\
  }\textbf {\bibinfo {volume} {115}},\ \bibinfo {pages} {050501} (\bibinfo
  {year} {2015})}\BibitemShut {NoStop}%
\bibitem [{\citenamefont {Skrzypczyk}\ \emph {et~al.}(2014)\citenamefont
  {Skrzypczyk}, \citenamefont {Short},\ and\ \citenamefont
  {Popescu}}]{skrzypczykWorkExtractionThermodynamics2014}%
  \BibitemOpen
  \bibfield  {author} {\bibinfo {author} {\bibfnamefont {P.}~\bibnamefont
  {Skrzypczyk}}, \bibinfo {author} {\bibfnamefont {A.~J.}\ \bibnamefont
  {Short}},\ and\ \bibinfo {author} {\bibfnamefont {S.}~\bibnamefont
  {Popescu}},\ }\bibfield  {title} {\bibinfo {title} {Work extraction and
  thermodynamics for individual quantum systems},\ }\href
  {https://doi.org/10.1038/ncomms5185} {\bibfield  {journal} {\bibinfo
  {journal} {Nature Communications}\ }\textbf {\bibinfo {volume} {5}},\
  \bibinfo {pages} {4185} (\bibinfo {year} {2014})}\BibitemShut {NoStop}%
\bibitem [{\citenamefont {Parrondo}\ \emph {et~al.}(2015)\citenamefont
  {Parrondo}, \citenamefont {Horowitz},\ and\ \citenamefont
  {Sagawa}}]{parrondoThermodynamicsInformation2015}%
  \BibitemOpen
  \bibfield  {author} {\bibinfo {author} {\bibfnamefont {J.~M.~R.}\
  \bibnamefont {Parrondo}}, \bibinfo {author} {\bibfnamefont {J.~M.}\
  \bibnamefont {Horowitz}},\ and\ \bibinfo {author} {\bibfnamefont
  {T.}~\bibnamefont {Sagawa}},\ }\bibfield  {title} {\bibinfo {title}
  {Thermodynamics of information},\ }\href {https://doi.org/10.1038/nphys3230}
  {\bibfield  {journal} {\bibinfo  {journal} {Nature Physics}\ }\textbf
  {\bibinfo {volume} {11}},\ \bibinfo {pages} {131} (\bibinfo {year}
  {2015})}\BibitemShut {NoStop}%
\bibitem [{\citenamefont {Micadei}\ \emph {et~al.}(2019)\citenamefont
  {Micadei}, \citenamefont {Peterson}, \citenamefont {Souza}, \citenamefont
  {Sarthour}, \citenamefont {Oliveira}, \citenamefont {Landi}, \citenamefont
  {Batalh{\~a}o}, \citenamefont {Serra},\ and\ \citenamefont
  {Lutz}}]{micadeiReversingDirectionHeat2019}%
  \BibitemOpen
  \bibfield  {author} {\bibinfo {author} {\bibfnamefont {K.}~\bibnamefont
  {Micadei}}, \bibinfo {author} {\bibfnamefont {J.~P.~S.}\ \bibnamefont
  {Peterson}}, \bibinfo {author} {\bibfnamefont {A.~M.}\ \bibnamefont {Souza}},
  \bibinfo {author} {\bibfnamefont {R.~S.}\ \bibnamefont {Sarthour}}, \bibinfo
  {author} {\bibfnamefont {I.~S.}\ \bibnamefont {Oliveira}}, \bibinfo {author}
  {\bibfnamefont {G.~T.}\ \bibnamefont {Landi}}, \bibinfo {author}
  {\bibfnamefont {T.~B.}\ \bibnamefont {Batalh{\~a}o}}, \bibinfo {author}
  {\bibfnamefont {R.~M.}\ \bibnamefont {Serra}},\ and\ \bibinfo {author}
  {\bibfnamefont {E.}~\bibnamefont {Lutz}},\ }\bibfield  {title} {\bibinfo
  {title} {Reversing the direction of heat flow using quantum correlations},\
  }\href {https://doi.org/10.1038/s41467-019-10333-7} {\bibfield  {journal}
  {\bibinfo  {journal} {Nature Communications}\ }\textbf {\bibinfo {volume}
  {10}},\ \bibinfo {pages} {2456} (\bibinfo {year} {2019})}\BibitemShut
  {NoStop}%
\bibitem [{\citenamefont {Kraus}\ and\ \citenamefont
  {Cirac}(2001)}]{krausOptimalCreationEntanglement2001}%
  \BibitemOpen
  \bibfield  {author} {\bibinfo {author} {\bibfnamefont {B.}~\bibnamefont
  {Kraus}}\ and\ \bibinfo {author} {\bibfnamefont {J.~I.}\ \bibnamefont
  {Cirac}},\ }\bibfield  {title} {\bibinfo {title} {Optimal creation of
  entanglement using a two-qubit gate},\ }\href
  {https://doi.org/10.1103/PhysRevA.63.062309} {\bibfield  {journal} {\bibinfo
  {journal} {Physical Review A}\ }\textbf {\bibinfo {volume} {63}},\ \bibinfo
  {pages} {062309} (\bibinfo {year} {2001})}\BibitemShut {NoStop}%
\bibitem [{\citenamefont {Merhav}\ and\ \citenamefont
  {Kafri}(2010)}]{merhavStatisticalPropertiesEntropy2010}%
  \BibitemOpen
  \bibfield  {author} {\bibinfo {author} {\bibfnamefont {N.}~\bibnamefont
  {Merhav}}\ and\ \bibinfo {author} {\bibfnamefont {Y.}~\bibnamefont {Kafri}},\
  }\bibfield  {title} {\bibinfo {title} {Statistical properties of entropy
  production derived from fluctuation theorems},\ }\href
  {https://doi.org/10.1088/1742-5468/2010/12/P12022} {\bibfield  {journal}
  {\bibinfo  {journal} {Journal of Statistical Mechanics: Theory and
  Experiment}\ }\textbf {\bibinfo {volume} {2010}},\ \bibinfo {pages} {P12022}
  (\bibinfo {year} {2010})}\BibitemShut {NoStop}%
\bibitem [{\citenamefont {Deffner}\ and\ \citenamefont
  {Lutz}(2011)}]{deffnerNonequilibriumEntropyProduction2011}%
  \BibitemOpen
  \bibfield  {author} {\bibinfo {author} {\bibfnamefont {S.}~\bibnamefont
  {Deffner}}\ and\ \bibinfo {author} {\bibfnamefont {E.}~\bibnamefont {Lutz}},\
  }\bibfield  {title} {\bibinfo {title} {Nonequilibrium {{Entropy Production}}
  for {{Open Quantum Systems}}},\ }\href
  {https://doi.org/10.1103/PhysRevLett.107.140404} {\bibfield  {journal}
  {\bibinfo  {journal} {Physical Review Letters}\ }\textbf {\bibinfo {volume}
  {107}},\ \bibinfo {pages} {140404} (\bibinfo {year} {2011})}\BibitemShut
  {NoStop}%
\bibitem [{\citenamefont {Baumgratz}\ \emph {et~al.}(2014)\citenamefont
  {Baumgratz}, \citenamefont {Cramer},\ and\ \citenamefont
  {Plenio}}]{baumgratzQuantifyingCoherence2014}%
  \BibitemOpen
  \bibfield  {author} {\bibinfo {author} {\bibfnamefont {T.}~\bibnamefont
  {Baumgratz}}, \bibinfo {author} {\bibfnamefont {M.}~\bibnamefont {Cramer}},\
  and\ \bibinfo {author} {\bibfnamefont {M.~B.}\ \bibnamefont {Plenio}},\
  }\bibfield  {title} {\bibinfo {title} {Quantifying {{Coherence}}},\ }\href
  {https://doi.org/10.1103/PhysRevLett.113.140401} {\bibfield  {journal}
  {\bibinfo  {journal} {Physical Review Letters}\ }\textbf {\bibinfo {volume}
  {113}},\ \bibinfo {pages} {140401} (\bibinfo {year} {2014})}\BibitemShut
  {NoStop}%
\bibitem [{\citenamefont {Aw}\ \emph {et~al.}(2021)\citenamefont {Aw},
  \citenamefont {Buscemi},\ and\ \citenamefont
  {Scarani}}]{awFluctuationTheoremsRetrodiction2021}%
  \BibitemOpen
  \bibfield  {author} {\bibinfo {author} {\bibfnamefont {C.~C.}\ \bibnamefont
  {Aw}}, \bibinfo {author} {\bibfnamefont {F.}~\bibnamefont {Buscemi}},\ and\
  \bibinfo {author} {\bibfnamefont {V.}~\bibnamefont {Scarani}},\ }\bibfield
  {title} {\bibinfo {title} {Fluctuation {{Theorems}} with {{Retrodiction}}
  rather than {{Reverse Processes}}},\ }\href
  {https://doi.org/10.1116/5.0060893} {\bibfield  {journal} {\bibinfo
  {journal} {AVS Quantum Science}\ }\textbf {\bibinfo {volume} {3}},\ \bibinfo
  {pages} {045601} (\bibinfo {year} {2021})},\ \Eprint
  {https://arxiv.org/abs/2106.08589} {arXiv:2106.08589} \BibitemShut {NoStop}%
\bibitem [{\citenamefont {Buscemi}\ and\ \citenamefont
  {Scarani}(2021)}]{buscemiFluctuationTheoremsBayesian2021}%
  \BibitemOpen
  \bibfield  {author} {\bibinfo {author} {\bibfnamefont {F.}~\bibnamefont
  {Buscemi}}\ and\ \bibinfo {author} {\bibfnamefont {V.}~\bibnamefont
  {Scarani}},\ }\bibfield  {title} {\bibinfo {title} {Fluctuation theorems from
  {{Bayesian}} retrodiction},\ }\href
  {https://doi.org/10.1103/PhysRevE.103.052111} {\bibfield  {journal} {\bibinfo
   {journal} {Physical Review E}\ }\textbf {\bibinfo {volume} {103}},\ \bibinfo
  {pages} {052111} (\bibinfo {year} {2021})}\BibitemShut {NoStop}%
\bibitem [{\citenamefont
  {Yunger~Halpern}(2017)}]{yungerhalpernJarzynskilikeEqualityOutoftimeordered2017}%
  \BibitemOpen
  \bibfield  {author} {\bibinfo {author} {\bibfnamefont {N.}~\bibnamefont
  {Yunger~Halpern}},\ }\bibfield  {title} {\bibinfo {title} {Jarzynski-like
  equality for the out-of-time-ordered correlator},\ }\href
  {https://doi.org/10.1103/PhysRevA.95.012120} {\bibfield  {journal} {\bibinfo
  {journal} {Physical Review A}\ }\textbf {\bibinfo {volume} {95}},\ \bibinfo
  {pages} {012120} (\bibinfo {year} {2017})}\BibitemShut {NoStop}%
\bibitem [{\citenamefont {Kwon}\ and\ \citenamefont
  {Kim}(2019)}]{kwonFluctuationTheoremsQuantum2019}%
  \BibitemOpen
  \bibfield  {author} {\bibinfo {author} {\bibfnamefont {H.}~\bibnamefont
  {Kwon}}\ and\ \bibinfo {author} {\bibfnamefont {M.~S.}\ \bibnamefont {Kim}},\
  }\bibfield  {title} {\bibinfo {title} {Fluctuation {{Theorems}} for a
  {{Quantum Channel}}},\ }\href {https://doi.org/10.1103/PhysRevX.9.031029}
  {\bibfield  {journal} {\bibinfo  {journal} {Physical Review X}\ }\textbf
  {\bibinfo {volume} {9}},\ \bibinfo {pages} {031029} (\bibinfo {year}
  {2019})}\BibitemShut {NoStop}%
\bibitem [{\citenamefont {Levy}\ and\ \citenamefont
  {Lostaglio}(2020)}]{levyQuasiprobabilityDistributionHeat2020}%
  \BibitemOpen
  \bibfield  {author} {\bibinfo {author} {\bibfnamefont {A.}~\bibnamefont
  {Levy}}\ and\ \bibinfo {author} {\bibfnamefont {M.}~\bibnamefont
  {Lostaglio}},\ }\bibfield  {title} {\bibinfo {title} {Quasiprobability
  {{Distribution}} for {{Heat Fluctuations}} in the {{Quantum Regime}}},\
  }\href {https://doi.org/10.1103/PRXQuantum.1.010309} {\bibfield  {journal}
  {\bibinfo  {journal} {PRX Quantum}\ }\textbf {\bibinfo {volume} {1}},\
  \bibinfo {pages} {010309} (\bibinfo {year} {2020})}\BibitemShut {NoStop}%
\bibitem [{\citenamefont {Park}\ \emph {et~al.}(2017)\citenamefont {Park},
  \citenamefont {Kim},\ and\ \citenamefont
  {Vedral}}]{parkFluctuationTheoremArbitrary2017}%
  \BibitemOpen
  \bibfield  {author} {\bibinfo {author} {\bibfnamefont {J.~J.}\ \bibnamefont
  {Park}}, \bibinfo {author} {\bibfnamefont {S.~W.}\ \bibnamefont {Kim}},\ and\
  \bibinfo {author} {\bibfnamefont {V.}~\bibnamefont {Vedral}},\ }\bibfield
  {title} {\bibinfo {title} {Fluctuation {{Theorem}} for {{Arbitrary Quantum
  Bipartite Systems}}},\ }\href@noop {} {\bibfield  {journal} {\bibinfo
  {journal} {arXiv:1705.01750 [quant-ph]}\ } (\bibinfo {year} {2017})},\
  \Eprint {https://arxiv.org/abs/1705.01750} {arXiv:1705.01750 [quant-ph]}
  \BibitemShut {NoStop}%
\bibitem [{\citenamefont {Micadei}\ \emph {et~al.}(2020)\citenamefont
  {Micadei}, \citenamefont {Landi},\ and\ \citenamefont
  {Lutz}}]{micadeiQuantumFluctuationTheorems2020}%
  \BibitemOpen
  \bibfield  {author} {\bibinfo {author} {\bibfnamefont {K.}~\bibnamefont
  {Micadei}}, \bibinfo {author} {\bibfnamefont {G.~T.}\ \bibnamefont {Landi}},\
  and\ \bibinfo {author} {\bibfnamefont {E.}~\bibnamefont {Lutz}},\ }\bibfield
  {title} {\bibinfo {title} {Quantum {{Fluctuation Theorems}} beyond
  {{Two-Point Measurements}}},\ }\href
  {https://doi.org/10.1103/PhysRevLett.124.090602} {\bibfield  {journal}
  {\bibinfo  {journal} {Physical Review Letters}\ }\textbf {\bibinfo {volume}
  {124}},\ \bibinfo {pages} {090602} (\bibinfo {year} {2020})}\BibitemShut
  {NoStop}%
\bibitem [{\citenamefont {Park}\ \emph {et~al.}(2020)\citenamefont {Park},
  \citenamefont {Nha}, \citenamefont {Kim},\ and\ \citenamefont
  {Vedral}}]{parkInformationFluctuationTheorem2020}%
  \BibitemOpen
  \bibfield  {author} {\bibinfo {author} {\bibfnamefont {J.~J.}\ \bibnamefont
  {Park}}, \bibinfo {author} {\bibfnamefont {H.}~\bibnamefont {Nha}}, \bibinfo
  {author} {\bibfnamefont {S.~W.}\ \bibnamefont {Kim}},\ and\ \bibinfo {author}
  {\bibfnamefont {V.}~\bibnamefont {Vedral}},\ }\bibfield  {title} {\bibinfo
  {title} {Information fluctuation theorem for an open quantum bipartite
  system},\ }\href {https://doi.org/10.1103/PhysRevE.101.052128} {\bibfield
  {journal} {\bibinfo  {journal} {Physical Review E}\ }\textbf {\bibinfo
  {volume} {101}},\ \bibinfo {pages} {052128} (\bibinfo {year}
  {2020})}\BibitemShut {NoStop}%
\bibitem [{\citenamefont {Dorner}\ \emph {et~al.}(2013)\citenamefont {Dorner},
  \citenamefont {Clark}, \citenamefont {Heaney}, \citenamefont {Fazio},
  \citenamefont {Goold},\ and\ \citenamefont
  {Vedral}}]{dornerExtractingQuantumWork2013}%
  \BibitemOpen
  \bibfield  {author} {\bibinfo {author} {\bibfnamefont {R.}~\bibnamefont
  {Dorner}}, \bibinfo {author} {\bibfnamefont {S.~R.}\ \bibnamefont {Clark}},
  \bibinfo {author} {\bibfnamefont {L.}~\bibnamefont {Heaney}}, \bibinfo
  {author} {\bibfnamefont {R.}~\bibnamefont {Fazio}}, \bibinfo {author}
  {\bibfnamefont {J.}~\bibnamefont {Goold}},\ and\ \bibinfo {author}
  {\bibfnamefont {V.}~\bibnamefont {Vedral}},\ }\bibfield  {title} {\bibinfo
  {title} {Extracting {{Quantum Work Statistics}} and {{Fluctuation Theorems}}
  by {{Single-Qubit Interferometry}}},\ }\href
  {https://doi.org/10.1103/PhysRevLett.110.230601} {\bibfield  {journal}
  {\bibinfo  {journal} {Physical Review Letters}\ }\textbf {\bibinfo {volume}
  {110}},\ \bibinfo {pages} {230601} (\bibinfo {year} {2013})}\BibitemShut
  {NoStop}%
\bibitem [{\citenamefont {Batalh{\~a}o}\ \emph {et~al.}(2014)\citenamefont
  {Batalh{\~a}o}, \citenamefont {Souza}, \citenamefont {Mazzola}, \citenamefont
  {Auccaise}, \citenamefont {Sarthour}, \citenamefont {Oliveira}, \citenamefont
  {Goold}, \citenamefont {De~Chiara}, \citenamefont {Paternostro},\ and\
  \citenamefont {Serra}}]{batalhaoExperimentalReconstructionWork2014}%
  \BibitemOpen
  \bibfield  {author} {\bibinfo {author} {\bibfnamefont {T.~B.}\ \bibnamefont
  {Batalh{\~a}o}}, \bibinfo {author} {\bibfnamefont {A.~M.}\ \bibnamefont
  {Souza}}, \bibinfo {author} {\bibfnamefont {L.}~\bibnamefont {Mazzola}},
  \bibinfo {author} {\bibfnamefont {R.}~\bibnamefont {Auccaise}}, \bibinfo
  {author} {\bibfnamefont {R.~S.}\ \bibnamefont {Sarthour}}, \bibinfo {author}
  {\bibfnamefont {I.~S.}\ \bibnamefont {Oliveira}}, \bibinfo {author}
  {\bibfnamefont {J.}~\bibnamefont {Goold}}, \bibinfo {author} {\bibfnamefont
  {G.}~\bibnamefont {De~Chiara}}, \bibinfo {author} {\bibfnamefont
  {M.}~\bibnamefont {Paternostro}},\ and\ \bibinfo {author} {\bibfnamefont
  {R.~M.}\ \bibnamefont {Serra}},\ }\bibfield  {title} {\bibinfo {title}
  {Experimental {{Reconstruction}} of {{Work Distribution}} and {{Study}} of
  {{Fluctuation Relations}} in a {{Closed Quantum System}}},\ }\href
  {https://doi.org/10.1103/PhysRevLett.113.140601} {\bibfield  {journal}
  {\bibinfo  {journal} {Physical Review Letters}\ }\textbf {\bibinfo {volume}
  {113}},\ \bibinfo {pages} {140601} (\bibinfo {year} {2014})}\BibitemShut
  {NoStop}%
\bibitem [{\citenamefont {Cerisola}\ \emph {et~al.}(2017)\citenamefont
  {Cerisola}, \citenamefont {Margalit}, \citenamefont {Machluf}, \citenamefont
  {Roncaglia}, \citenamefont {Paz},\ and\ \citenamefont
  {Folman}}]{cerisolaUsingQuantumWork2017}%
  \BibitemOpen
  \bibfield  {author} {\bibinfo {author} {\bibfnamefont {F.}~\bibnamefont
  {Cerisola}}, \bibinfo {author} {\bibfnamefont {Y.}~\bibnamefont {Margalit}},
  \bibinfo {author} {\bibfnamefont {S.}~\bibnamefont {Machluf}}, \bibinfo
  {author} {\bibfnamefont {A.~J.}\ \bibnamefont {Roncaglia}}, \bibinfo {author}
  {\bibfnamefont {J.~P.}\ \bibnamefont {Paz}},\ and\ \bibinfo {author}
  {\bibfnamefont {R.}~\bibnamefont {Folman}},\ }\bibfield  {title} {\bibinfo
  {title} {Using a quantum work meter to test non-equilibrium fluctuation
  theorems},\ }\href {https://doi.org/10.1038/s41467-017-01308-7} {\bibfield
  {journal} {\bibinfo  {journal} {Nature Communications}\ }\textbf {\bibinfo
  {volume} {8}},\ \bibinfo {pages} {1241} (\bibinfo {year} {2017})}\BibitemShut
  {NoStop}%
\bibitem [{\citenamefont {An}\ \emph {et~al.}(2015)\citenamefont {An},
  \citenamefont {Zhang}, \citenamefont {Um}, \citenamefont {Lv}, \citenamefont
  {Lu}, \citenamefont {Zhang}, \citenamefont {Yin}, \citenamefont {Quan},\ and\
  \citenamefont {Kim}}]{anExperimentalTestQuantum2015}%
  \BibitemOpen
  \bibfield  {author} {\bibinfo {author} {\bibfnamefont {S.}~\bibnamefont
  {An}}, \bibinfo {author} {\bibfnamefont {J.-N.}\ \bibnamefont {Zhang}},
  \bibinfo {author} {\bibfnamefont {M.}~\bibnamefont {Um}}, \bibinfo {author}
  {\bibfnamefont {D.}~\bibnamefont {Lv}}, \bibinfo {author} {\bibfnamefont
  {Y.}~\bibnamefont {Lu}}, \bibinfo {author} {\bibfnamefont {J.}~\bibnamefont
  {Zhang}}, \bibinfo {author} {\bibfnamefont {Z.-Q.}\ \bibnamefont {Yin}},
  \bibinfo {author} {\bibfnamefont {H.~T.}\ \bibnamefont {Quan}},\ and\
  \bibinfo {author} {\bibfnamefont {K.}~\bibnamefont {Kim}},\ }\bibfield
  {title} {\bibinfo {title} {Experimental test of the quantum {{Jarzynski}}
  equality with a trapped-ion system},\ }\href
  {https://doi.org/10.1038/nphys3197} {\bibfield  {journal} {\bibinfo
  {journal} {Nature Physics}\ }\textbf {\bibinfo {volume} {11}},\ \bibinfo
  {pages} {193} (\bibinfo {year} {2015})}\BibitemShut {NoStop}%
\bibitem [{\citenamefont {Masuyama}\ \emph {et~al.}(2018)\citenamefont
  {Masuyama}, \citenamefont {Funo}, \citenamefont {Murashita}, \citenamefont
  {Noguchi}, \citenamefont {Kono}, \citenamefont {Tabuchi}, \citenamefont
  {Yamazaki}, \citenamefont {Ueda},\ and\ \citenamefont
  {Nakamura}}]{masuyamaInformationtoworkConversionMaxwell2018}%
  \BibitemOpen
  \bibfield  {author} {\bibinfo {author} {\bibfnamefont {Y.}~\bibnamefont
  {Masuyama}}, \bibinfo {author} {\bibfnamefont {K.}~\bibnamefont {Funo}},
  \bibinfo {author} {\bibfnamefont {Y.}~\bibnamefont {Murashita}}, \bibinfo
  {author} {\bibfnamefont {A.}~\bibnamefont {Noguchi}}, \bibinfo {author}
  {\bibfnamefont {S.}~\bibnamefont {Kono}}, \bibinfo {author} {\bibfnamefont
  {Y.}~\bibnamefont {Tabuchi}}, \bibinfo {author} {\bibfnamefont
  {R.}~\bibnamefont {Yamazaki}}, \bibinfo {author} {\bibfnamefont
  {M.}~\bibnamefont {Ueda}},\ and\ \bibinfo {author} {\bibfnamefont
  {Y.}~\bibnamefont {Nakamura}},\ }\bibfield  {title} {\bibinfo {title}
  {Information-to-work conversion by {{Maxwell}}'s demon in a superconducting
  circuit quantum electrodynamical system},\ }\href
  {https://doi.org/10.1038/s41467-018-03686-y} {\bibfield  {journal} {\bibinfo
  {journal} {Nature Communications}\ }\textbf {\bibinfo {volume} {9}},\
  \bibinfo {pages} {1291} (\bibinfo {year} {2018})}\BibitemShut {NoStop}%
\bibitem [{\citenamefont {Solfanelli}\ \emph {et~al.}(2021)\citenamefont
  {Solfanelli}, \citenamefont {Santini},\ and\ \citenamefont
  {Campisi}}]{solfanelliExperimentalVerificationFluctuation2021}%
  \BibitemOpen
  \bibfield  {author} {\bibinfo {author} {\bibfnamefont {A.}~\bibnamefont
  {Solfanelli}}, \bibinfo {author} {\bibfnamefont {A.}~\bibnamefont
  {Santini}},\ and\ \bibinfo {author} {\bibfnamefont {M.}~\bibnamefont
  {Campisi}},\ }\bibfield  {title} {\bibinfo {title} {Experimental
  {{Verification}} of {{Fluctuation Relations}} with a {{Quantum Computer}}},\
  }\href {https://doi.org/10.1103/PRXQuantum.2.030353} {\bibfield  {journal}
  {\bibinfo  {journal} {PRX Quantum}\ }\textbf {\bibinfo {volume} {2}},\
  \bibinfo {pages} {030353} (\bibinfo {year} {2021})}\BibitemShut {NoStop}%
\bibitem [{IBM(2022)}]{IBM}%
  \BibitemOpen
  \bibfield  {title} {\bibinfo {title} {{IBM Quantum}},\ }\href@noop {}
  {\bibfield  {journal} {\bibinfo  {journal}
  {\url{https://quantum-computing.ibm.com/}}\ } (\bibinfo {year}
  {2022})}\BibitemShut {NoStop}%
\bibitem [{\citenamefont {Vatan}\ and\ \citenamefont
  {Williams}(2004)}]{vatanOptimalQuantumCircuits2004}%
  \BibitemOpen
  \bibfield  {author} {\bibinfo {author} {\bibfnamefont {F.}~\bibnamefont
  {Vatan}}\ and\ \bibinfo {author} {\bibfnamefont {C.}~\bibnamefont
  {Williams}},\ }\bibfield  {title} {\bibinfo {title} {Optimal quantum circuits
  for general two-qubit gates},\ }\href
  {https://doi.org/10.1103/PhysRevA.69.032315} {\bibfield  {journal} {\bibinfo
  {journal} {Physical Review A}\ }\textbf {\bibinfo {volume} {69}},\ \bibinfo
  {pages} {032315} (\bibinfo {year} {2004})}\BibitemShut {NoStop}%
\bibitem [{\citenamefont {Herrera}\ \emph {et~al.}(2021)\citenamefont
  {Herrera}, \citenamefont {Peterson}, \citenamefont {Serra},\ and\
  \citenamefont {D'Amico}}]{herreraEasyAccessEnergy2021}%
  \BibitemOpen
  \bibfield  {author} {\bibinfo {author} {\bibfnamefont {M.}~\bibnamefont
  {Herrera}}, \bibinfo {author} {\bibfnamefont {J.~P.~S.}\ \bibnamefont
  {Peterson}}, \bibinfo {author} {\bibfnamefont {R.~M.}\ \bibnamefont
  {Serra}},\ and\ \bibinfo {author} {\bibfnamefont {I.}~\bibnamefont
  {D'Amico}},\ }\bibfield  {title} {\bibinfo {title} {Easy {{Access}} to
  {{Energy Fluctuations}} in {{Nonequilibrium Quantum Many-Body Systems}}},\
  }\href {https://doi.org/10.1103/PhysRevLett.127.030602} {\bibfield  {journal}
  {\bibinfo  {journal} {Physical Review Letters}\ }\textbf {\bibinfo {volume}
  {127}},\ \bibinfo {pages} {030602} (\bibinfo {year} {2021})}\BibitemShut
  {NoStop}%
\end{thebibliography}

\providecommand{\noopsort}[1]{}\providecommand{\singleletter}[1]{#1}%

\end{document}